\documentclass[useAMS,usenatbib]{mn2e}

\usepackage{graphicx}
\DeclareGraphicsExtensions{.eps,.jpg,.pdf,.mps,.png,.gif}
\usepackage{subfigure}

\usepackage{xspace}
\usepackage{amsmath, amsfonts, amssymb}
\usepackage{gensymb}
\usepackage{hhline}
\usepackage{lscape}

\def\reff@jnl#1{{\rm#1\/}}
\def\aj{\reff@jnl{AJ}}                  
\def\araa{\reff@jnl{ARA\&A}}            
\def\apj{\reff@jnl{ApJ}}                
\def\apjl{\reff@jnl{ApJ}}               
\def\apjs{\reff@jnl{ApJS}}              
\def\ao{\reff@jnl{Appl.Optics}}         
\def\apss{\reff@jnl{Ap\&SS}}            
\def\aap{\reff@jnl{A\&A}}               
\def\aapr{\reff@jnl{A\&A~Rev.}}         
\def\aaps{\reff@jnl{A\&AS}}             
\def\baas{\reff@jnl{BAAS}}              
\def\jrasc{\reff@jnl{JRASC}}            
\def\memras{\reff@jnl{MmRAS}}           
\def\mnras{\reff@jnl{MNRAS}}            
\def\pra{\reff@jnl{Phys.Rev.A}}         
\def\prb{\reff@jnl{Phys.Rev.B}}         
\def\prc{\reff@jnl{Phys.Rev.C}}         
\def\prd{\reff@jnl{Phys.Rev.D}}         
\def\prl{\reff@jnl{Phys.Rev.Lett}}      
\def\pasp{\reff@jnl{PASP}}              
\def\pasj{\reff@jnl{PASJ}}              
\def\qjras{\reff@jnl{QJRAS}}            
\def\skytel{\reff@jnl{S\&T}}            
\def\solphys{\reff@jnl{Solar~Phys.}}    
\def\sovast{\reff@jnl{Soviet~Ast.}}     
\def\ssr{\reff@jnl{Space~Sci.Rev.}}     
\def\zap{\reff@jnl{ZAp}}                
\def\nat{\reff@jnl{Nature}}             
\def\nar{\reff@jnl{NewA Rev.}}  	
\def\na{\reff@jnl{NewA}}  		

\newcommand{\fref}{Fig.~\ref}
\newcommand{\Fref}{Fig.~\ref}

\newcommand{\Eref}{Equation~\ref}
\newcommand{\tref}{Tab.~\ref}
\newcommand{\Tref}{Tab.~\ref}
\newcommand{\sref}{section~\ref}

\newcommand{\Aref}{Appendix~\ref}
\newcommand{\ie}{i.e.\xspace}

\newcommand{\galfit}{\textsc{Galfit}\xspace}
\newcommand{\tinytim}{\textsc{TinyTim}\xspace}

\newcommand{\be}{\begin{eqnarray}}
\newcommand{\ee}{\end{eqnarray}}

\title[The Role of Secular Evolution in NLS1 Galaxies]{The Role of Secular Evolution in the Black Hole Growth of Narrow-Line Seyfert 1 Galaxies}

\author[G. Orban de Xivry et al.]
{G. Orban de Xivry$^{1}$\thanks{E-mail: xivry@mpe.mpg.de}, R. Davies$^{1}$, M. Schartmann$^{1,2}$, S. Komossa$^{1,3,4,5}$,  A. Marconi$^{6}$
\newauthor E. Hicks$^{7}$, H. Engel$^{1}$, L. Tacconi$^{1}$\\
$^{1}$Max Planck Institut f\"ur extraterrestrische Physik, Giessenbachstrasse, D-85748 Garching, Germany\\
$^{2}$Universit\"ats-Sternwarte M\"unchen, Scheinerstrasse 1, D-81679 M\"unchen, Germany \\
$^{3}$Technische Universit\"at M\"unchen, Fakult\"at f\"ur Physik, James-Franck-Strasse, 85748 Garching, Germany  \\
$^{4}$Excellence Cluster Universe, TUM, Boltzmannstrasse 2, 85748 Garching, Germany \\
$^{5}$Max Planck Institut f\"ur Plasmaphysik, Boltzmannstrasse 2, 85748 Garching, Germany \\
$^{6}$Dipartimento di Fisica e Astronomia, Universit\'a degli Studi di Firenze, Largo E. Fermi 2, 50125 Firenze, Italy\\ 
$^{7}$Department of Astronomy, University of Washington, Box 351580, Seattle, WA 98195, USA
}

\begin{document}

\date{\today}

\pagerange{\pageref{firstpage}--\pageref{lastpage}} \pubyear{2011}

\maketitle

\label{firstpage}


\begin{abstract}
Narrow-Line Seyfert 1 (NLS1) galaxies show extreme properties with respect to the other Seyfert galaxies. Indeed, they are thought to be 
accreting at Eddington rates and to possess low mass black holes. Therefore, they may represent a key class of objects for understanding the co-evolution of black holes and their host galaxies.
We propose that NLS1s represent a class of AGN in which the black hole growth is, and has always been, dominated by secular evolution. 
Firstly, by looking at the NLS1 host galaxy properties in the literature, we show that the evolution of NLS1s is presently driven by secular processes, much more so than for Broad-Line Seyfert 1s (BLS1s). 
Secondly, we study the bulges of NLS1 and BLS1 galaxies. Our results demonstrate that NLS1 host bulges are pseudo-bulges and are statistically different from BLS1 bulges. This difference points to the particular importance of secular processes in the past evolution of their hosts.
We build on this result to understand the implications on their evolution and the duration of their duty cycle. We show that NLS1s are not necessarily in a special phase of black hole growth and that several Gyr are required for their black hole masses to become similar to BLS1s. Finally, in the light of our results, we discuss the location of NLS1 galaxies on the M$_{\mathrm{BH}}$-$\sigma$ plane and speculate about the connection between the NLS1 galaxy properties and their black hole spin.
\end{abstract}



\section{Introduction}
Since their discovery, Narrow-Line Seyfert 1 (NLS1) galaxies have always been recognized as particular objects holding important clues on the driving mechanisms of nuclear activity. First identified as objects similar to Seyfert 1s with narrower Balmer lines, they were soon  recognized as having exceptional spectral properties, both in their emission lines and in their continuum \citep[see][for a review]{K08}. 
In fact, as shown through principal components analysis  \citep[e.g.][]{B02}, NLS1 galaxies are  mostly clustered at one extreme end of the AGN parameter space (strongest FeII/[OIII] emission and lowest luminosity). 
Likewise, they are thought to possess small mass black holes and to have high Eddington accretion rates. In this sense, NLS1s could represent key objects in understanding the AGN phenomena and the  co-evolution of massive black holes and their host galaxies.

While the main defining criteria of NLS1s with respect to BLS1s is  the empirical threshold at FWHM(H$_{\beta}$) $\sim$ 2000 km s$^{-1}$, the properties of NLS1s have been extensively studied across many wavelength ranges. Trends and correlations have been identified using first small samples and later corroborated by larger surveys \citep[e.g.][]{VV01,W+02,Z+06}. 
Many scenarios have been considered to explain these properties, in particular their high accretion rates ($L/L_{\mathrm{Edd}} \simeq 1$, \citealp[e.g.][and reference therein]{B02,Gru04}) and low black hole masses (typically of order $10^6 M_{\sun}$, \citealp[e.g.][]{BBF96,Z+06,R+07}), but also, e.g., outflows, winds and density effects, high metallicity, particular broad line region thicknesses and densities, etc. \citep[see][and references therein]{K08}. 
While these scenarios can elucidate the nuclear properties of NLS1s, they hardly explain the origin of the fundamental differences between NLS1s and BLS1s and, in particular, that NLS1s appear to be more than just Seyfert 1s with narrow lines. 
A few key questions could be formulated as follows:
which particular mechanisms would lead to the Eddington accretion rates commonly seen in NLS1s but observed less often in BLS1s? 
What causes the difference in black hole growth of NLS1s and BLS1s that results in low mass black holes in the former case?
Could differing host galaxy evolution explain the differences between NLS1 and BLS1 galaxies?

In this paper, rather than studying the active nuclei, we investigate the host galaxies of NLS1s and contrast their properties to those of BLS1s, pursuing the hypothesis that different host galaxy evolution could explain the differences between NLS1s and BLS1s. 
In particular, we explore the relative role of secular processes in the evolution of NLS1 and BLS1 galaxies. 
 Reviewing the literature on the morphology and the star formation in  NLS1 and BLS1 hosts, we emphasize, in \sref{sec:host}, the present-day differences in their respective host galaxies.
Afterwards, in section \ref{sec:bulges}, we turn to the past evolution of NLS1 and BLS1 hosts. We perform a bulge-disk decomposition of samples of NLS1 and BLS1 galaxies and look at their bulge properties. Using previously established criteria \citep[][]{KK04,FD08,G09}, we are able to distinguish pseudo- from classical bulges. This enables us to determine the main processes that have driven the evolution of the NLS1 and BLS1 hosts. We analyse the differences of NLS1 and BLS1 host bulge property distributions, concluding that NLS1 galaxies contain pseudo-bulges and, hence, have always been dominated by secular evolution.
Finally, in \sref{sec:population} and \sref{sec:discussion}, we investigate the cosmological context of the NLS1 host phenomenon driven by such an evolutionary mode. 
 We then note the link between secular evolution and rapidly spinning black holes, and speculate on the location of NLS1 galaxies on the M$_{\mathrm{BH}} - \sigma$ relation.
We conclude by summarizing our picture of the NLS1 galaxy phenomenon and present ways to further test our proposition.

When calculating distances and look-back times, we assume a general relativistic Friedmann-Robertson-Walker (FRW) cosmology with matter-density parameter $\Omega_m=0.3$, vacuum energy-density parameter $\Omega_{\Lambda}=0.7$ and Hubble parameter $H_0 = 70$ km s$^{-1}$ Mpc$^{-1}$.\\


\section{Secular evolution in NLS1 host galaxies}\label{sec:host}
In this section, we review the literature that has been published concerning the differences in the host galaxy properties of NLS1s and BLS1s. We focus in particular on the morphology and the star formation rate, and emphasize the role of present secular processes in distinguishing between these two classes of type~1 AGN.

\subsection{Morphological properties}
\subsubsection{Large-scale bars}
The morphology of NLS1 host galaxies has been studied in several papers \citep[][]{CKG03,DCK06,O+07,R+07}.
The main results can be summarized as follows:

{\it NLS1 host galaxies are likely to be strongly barred (much more than BLS1 ones) and their nuclear dust morphology is likely to be a grand-design spiral.}

\begin{figure}
     \centering
     \includegraphics[width=8cm]{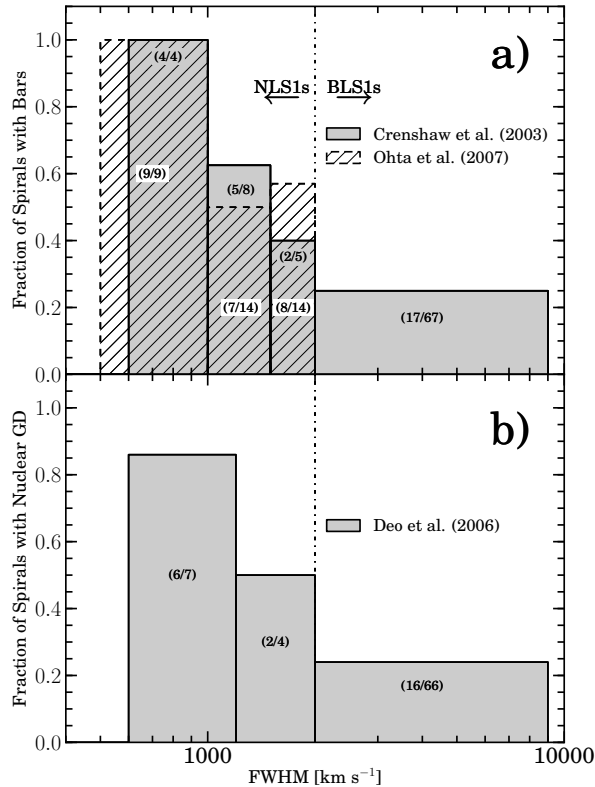}
     \caption{Histograms presenting the fraction of Seyfert 1 spirals a) with bars  and b) with nuclear grand-design spirals  as a function of the FWHM of their broad emission line, $H{\beta}$. Drawn from data reported in \citet{CKG03}, \citet{O+07} and \citet{DCK06}.}
     \label{fig:bar}
\end{figure}

The bar frequency among NLS1 and BLS1 host galaxies has been studied by \citet{CKG03} and \citet{O+07}.
The first paper reports a visual study based on an HST survey of 91 Seyfert galaxies (13 NLS1s and 78 BLS1s)  at $z\leq0.035$ \citep{MGT98} and 6 additional NLS1s at $z\leq0.084$ (HST archives in the \citealp{V+01} sample). The results are striking: 91\% of the sample is classified as spiral galaxies among which 65\% of NLS1s have bars, and 25\% of BLS1s have bars. More particularly, the authors look at the fraction of barred spiral galaxies in their sample as a function of the full width at half maximum (FWHM) of the broad component of the H$_{\beta}$ emission line.
As presented in \fref{fig:bar}, they obtain a clear difference between the two regimes of NLS1s and BLS1s.
We note that in this figure, since FWHM measurements are not available for every BLS1, we have made a single bin for the BLS1 class.

The results from \citet{O+07} are more conservative. They use an heterogeneous set of data of NLS1 galaxies and look at the morphology and the possible trends with the NLS1 properties.
They perform a visual  and a quantitative classification based on ellipse fitting of isophotes.
While they confirm the high bar fraction among (spiral) NLS1 hosts, they do not observe a clear trend  with the FWHM. Nevertheless, if we consider only the fraction of spirals with strong bars (SB but not SAB), the trend does appear clearly using their visual classification (as represented in \fref{fig:bar}).
Turning to their quantitative classification, we note that one of the criteria \citeauthor{O+07} use to identify the bars is an ellipticity $\epsilon_{\mathrm{bar}} \geq 0.25$, where $\epsilon_{\mathrm{bar}}=\mathrm{max}(\epsilon_{\mathrm{galaxy}})$. However a common practice to identify {\it strong} bars is to use $\epsilon_{\mathrm{bar}} \geq 0.45$ \citep[e.g.][]{SPK00}. 
Applying this latter criterion on their sample by examining the ellipse fit plots, we obtain, for the respective bins in \Fref{fig:bar} (\ie 500--1000, 1000--1500, 1500--2000 km s$^{-1}$), bar fractions (\ie 89\%, 46\% and 57\%) similar to their visual classification (\ie 100\%, 50\% and 57\%, see \Fref{fig:bar}).

\subsubsection{Circumnuclear morphology}
In a study of nuclear dust morphology in a matched-paired sample of active/inactive galaxies and barred/unbarred galaxies, \citet{Mar+03} show that grand-design nuclear dust spirals are only found in galaxies with a large-scale bar. 
However, while not finding any universal nuclear morphology in active galaxies, they do find similar features in the circumnuclear environments of both active and inactive galaxies.
In another study, \citet{DCK06} investigate the nuclear dust morphology in NLS1 and BLS1 host galaxies based only on the HST survey conducted by \citet{MGT98}. Their study also shows that the grand-design nuclear dust spirals are largely present in barred galaxies. They classify the nuclear structures and find that (1) the nuclear dust morphologies in NLS1/BLS1, Barred/Unbarred are mainly nuclear dust spirals, (2) in the ``nuclear dust spiral'' class, NLS1s are more likely to have grand-design spirals than BLS1s. 
\Fref{fig:bar} also shows that  this fraction of grand-design spirals in NLS1s follows the same trend with the FWHM(H$_{\beta}$)  as the bar fraction.

As we expect strong bars to drive a circumnuclear spiral structure \citep[][]{Mac04a,Mac04b} and to drive gas inwards \citep[][]{Sak+99,She+05}, we expect the presence of such asymmetries in host galaxies to result in an enhanced star formation in the central kiloparsecs.

\subsection{Star formation} 

In a recent paper, \citet{S+10} study the link between star formation in the central kpc vs. FWHM(H$_{\beta}$) in NLS1 and BLS1 host galaxies. After discussing carefully possible luminosity and distance effects, they conclude that NLS1s are associated with more intense star formation than  BLS1s (with on average a star formation to AGN ratio $>2$ times larger in NLS1s). More generally, they find that type 1 AGNs with narrower broad emission lines reside in hosts containing more intense star-forming regions.

Finally, they find a connection between high Eddington ratio and high star formation rates concluding that NLS1s are characterized by smaller black hole (BH) mass, larger Eddington ratio and stronger star formation activity compared to their broad-line counterparts.

\subsection{Secular processes all the way to the SMBH}
As discussed above, the current morphology of NLS1 host galaxies is distinguishable from other Seyfert galaxies.
Indeed, in contrast to BLS1s, NLS1 galaxies are likely to be  strongly barred and to show more intense central star formation.
This is in line with the fact that bars are known to drive gas into the central kiloparsecs \citep[][]{Sak+99,She+05}, and that nuclear star formation is enhanced in barred galaxies \citep{HFS97}.

While no universal fueling mechanism for low-luminosity AGNs seems to operate in galactic nuclei \citep{Mar+03}, the NLS1 host morphology typically exhibits a circumnuclear grand-design spiral.
This appears to be linked to the presence of strong bars \citep{Mar+03,DCK06}, and indeed bars are able to drive circumnuclear spiral structures \citep{Mac04a,Mac04b}.
Hence, NLS1 galaxies show uninterrupted asymmetries able to drive the gas inwards from a few kpc to a few tens of pc. The particular strength of secular processes in NLS1s could therefore account for the high central star formation and presumably to the large Eddington rates observed in NLS1s.

\section{Bulges of NLS1 host galaxies}\label{sec:bulges}
Since strong secular evolution is currently occurring in NLS1s, it is important to ask whether or not secular processes have shaped the NLS1 host galaxies by dominating their past evolution and hence influencing their black hole growth.

We address this issue by examining the bulges of NLS1  host galaxies and comparing them to those of BLS1s, since
one can expect to observe evolutionary dependent bulge characteristics.
Specifically, an evolution driven mainly by galaxy mergers will result in different bulge properties than if the evolution is mainly driven  by internal secular evolution.

In this section, we compare NLS1 and BLS1 galaxies   by performing a photometric bulge-disk decomposition of homogeneous samples of NLS1s and BLS1s. This comparison is put in perspective with previous studies on the distinction between pseudo- and classical bulges in inactive galaxies \citep[][]{KK04,FD08,G09}, and also on the bulge-disk decomposition of two galaxy samples composed exclusively of NLS1s \citep[][]{R+07,M+11}.

\subsection{Pseudo-bulges and secular evolution in disk galaxies}\label{ssec:pseudo}

\citet{KK04} review in detail the formation of pseudo-bulges by secular processes. As dense central components of galaxies, pseudo-bulges differ from classical bulges in that they were made slowly by disks out of disk material while classical bulges are ``merger-built'' bulges. Therefore, pseudo-bulges are formed by internal secular processes such as bar instabilities, spiral structures, etc. as opposed to galaxy mergers or external secular evolution (minor mergers, prolonged gas infall, etc.).

As pseudo-bulges retain memory of their disky origin, it is possible to disentangle them from classical bulges. \citet{KK04} pointed out the S\'ersic index as one way to identify  them. Indeed, since a pseudo-bulge  forms from gas accreting from the disk, it has a surface brightness profile similar to that of the outer disk and therefore would have a low S\'ersic index $n_b \sim 1-2$.

\citet{FD08} have studied in great detail the structure of classical bulges and pseudo-bulges using high-resolution data (77 inactive galaxies with data in the HST archive and $z\lesssim$0.01). They use morphological signatures to first visually classify the bulges of nearby galaxies as pseudo or classical.
They then perform a bulge-disk decomposition and study in particular the distribution of  S\'ersic indices. They find that, statistically, pseudo-bulges have S\'ersic indices $n_b < 2$ while classical bulges have $n_b > 2$.
This result shows  that the S\'ersic index is a good statistical tool to test if a class of objects has classical or pseudo-bulges. 

Finally, \citet{G09} also study pseudo- and classical bulges using a large, low resolution, SDSS sample  of galaxies ($\sim$1000 inactive galaxies with $0.02\leq z \leq 0.07$). He uses the position in the $<\mu_e> - r_e$ plane, also called the Kormendy relation\footnote{relation between the mean effective surface brightness within the effective radius $<\mu_e>$ and the half-light radius $r_e$,  which is a projection of the photometric fundamental plane.}, to study the bulge properties and identifies pseudo-bulges as being fainter in surface brightness for a given half-light radius (much fainter than predicted by the correlation fit to elliptical galaxies). Where \citet{G09} clearly sees independent groups in his $i$-band density plot of the Kormendy relation ($<\mu_e> - r_e$ relation), \citet{FD08,FD10} only find that pseudo-bulges scatter around the photometric projections of the fundamental plane.

Nevertheless, all these authors \citep{FD08,G09,FD10} agree that most of the pseudo-bulges have  a low S\'ersic index $n_b<2$  and that they tend to be less prominent than classical bulges (in particular, they tend to have a low bulge-to-total light ratio).

Based on these considerations for bulge classification in inactive galaxies, we will use  the S\'ersic index to identify the prevailing bulge type in the NLS1 and BLS1 populations as pseudo- or classical.
We will then study the prominence of NLS1 and BLS1 bulges.

\subsection{Bulge/disk decomposition}\label{sec:selection}

Building on the work of \citet{FD08} for inactive galaxies, we select archive HST images of Seyfert galaxies to study the bulges of active galaxies, in order to assess whether the bulge characteristics of NLS1s and BLS1s might explain, by their evolutionary implications, the distinctions between these two classes of AGN.
Crucially, by performing the bulge-disk decomposition for samples of both NLS1s and BLS1s, we  minimize the impact of any systematic errors that our fitting procedure might generate.

We select  NLS1 and BLS1 galaxies from the \citet{MGT98} HST imaging survey of nearby AGNs. This 
survey contains a uniform sample of 91 Seyfert 1 galaxies at $z \leq 0.035$ observed with the Wide Field Planetary Camera 2 (WFPC2) through the F606W filter. This sample, also used by \cite{CKG03} and \cite{DCK06} in their morphological studies, contains 11 NLS1 galaxies and 80 BLS1 galaxies. The 11 NLS1 galaxies are genuine NLS1s as identified by \citet{VVG01} on the basis of their optical spectra (broad component of $H{\beta} < 2000$km s$^{-1}$ and strong FeII emission) and are listed in the catalog of \cite{VV10}. We therefore initially select all the 11 NLS1 galaxies available, as well as 21 of the 80 BLS1s. 
The BLS1 sample selection is made in a way to roughly match the $\sim$25\% fraction of such hosts that are strongly barred \citep{CKG03}. 
Their individual Seyfert classifications are reported in \Tref{tab:samples}. These are taken from \citet{VV10} with 2 exceptions.
IC1816 is classified by \citet{M+04} as a type~1. Because their spectrum shows clear evidence of broad emission, we have adopted this classification.
For NGC5252, we have followed the classification as a type~1.9 given in \citet{OM93}, because the presence of broad H${\alpha}$, with a measured FWHM of $\sim$2500 km s$^{-1}$, is confirmed by \citet{AP+96}.
We note that the broad H$\alpha$ is very clear in polarized light \citep{T10}, and shows dramatic variations over a period of several years.
This may be why there is some uncertainty about its classification as a type~1 or~2. 
Following this initial selection, we refine our sample based on the limited field of view (FoV) of WFPC2 by rejecting objects with $z<0.010$ or scales $\le$0.23 kpc/\arcsec~in order to obtain a reasonable minimum FoV of $\gtrsim$8$\times$8kpc on each object. This criterion ensures that the disk of the host can be fitted properly, while the redshift limit of the original source catalogue ensures that the bulge is sufficiently well resolved.
The two final samples for which we performed the bulge-disk decomposition are composed of 10 NLS1s and 19 BLS1s, and are given in \Tref{tab:samples}. 
Finally, we check that the mean redshift is not strongly biased with respect to the NLS1 sample ($<z_{\mathrm{NLS1}}>=0.024$, $<z_{\mathrm{BLS1}}> = 0.027$), and that no particular circumnuclear morphology has been selected. 

\begin{table*}
\caption{HST sample of NLS1s and BLS1s. Columns: (1) object name; (2)-(4) J2000 coordinates and redshift from Nasa/IPAC Extragalactic database (NED); (5) luminosity distance in Mpc for an $H_0 = 70$km s$^{-1}$ Mpc$^{-1}$, $\Omega_m=0.3$ and $\Omega_{\Lambda} = 0.7$ cosmology; (6) respective scale, the WFPC2 pixel scale is 0.0456\arcsec; (7) morphological classification \citep[from][MGT]{MGT98}; (8) Seyfert classification according to the \citet{VV10} catalog except for IC1816 and NGC5252, see also the main text. S1 classification designates a type 1 AGN with unspecified sub-type; (9) FWHM of the broad component of $H{\beta}$ (or in a few cases, $H\alpha$), the NLS1 measurements are from \citet{VVG01}, the BLS1 measurements are taken from \citet{CKG03}, see also references therein.}
\label{tab:samples}
\begin{minipage}{140mm}
     \centering
     \begin{tabular}{lcrcrlccccr}
     \hline
     Object  Name    &    R.A. &   Dec.     & $z$ & $D$    &  px sc.         & Morpho.& AGN  & FWHM\\
	             &	(J2000)& (J2000)    &	  &  (Mpc) &  (kpc/\arcsec)  &  (MGT) & type & (km s$^{-1}$) \\
     \hline
     \hline
     \textsc{NLS1 sample} & & & & & & &\\
     \hline
     KUG1136 &  11 39 13.9 &  +33 55 51  & 0.032	& 131.84  & 0.64 &    SB0   & 	 S1n	& 1145  \\	
     MRK0042 &  11 53 41.8 &  +46 12 43  & 0.024	& 99.83   & 0.48 &    SBa   &  	 S1n	& 865   \\	  
     MRK0335\footnote{From the bulge disk decomposition, MRK335 seems better described by a unique high S\'ersic profile. This object is therefore rejected from our bulge analysis.} 
	     &  00 06 19.5 &  +20 12 11  & 0.025	& 109.12  & 0.50 &    ?     &  S1n   & 1350  \\  
     MRK0359 &  01 27 32.5 &  +19 10 44  & 0.017	& 71.31   & 0.35 &    SBb/c &   S1n  &  900   \\	  
     MRK0382 &  07 55 25.3 &  +39 11 10  & 0.034	& 139.74  & 0.68 &    SBa   &  S1n   & 1280  \\	  
     MRK0493 &  15 59 09.6 &  +35 01 47  & 0.031	& 127.87  & 0.62 &    S(B)a &   S1n  & 740   \\	  	  
     MRK0766 &  12 18 26.5 &  +29 48 47  & 0.012	& 50.65   & 0.25 &    SBc   &  S1n  & 1630  \\	  
     MRK0896 &  20 46 20.8 &  $-$02 48 45& 0.027	& 111.91  & 0.54 &    Sc    &   S1n & 1135  \\	  	  
     MRK1044	&  02 30 05.5 &  $-$08 59 53& 0.016	& 67.20   & 0.33 &    Sa    &  S1n  &  1010  \\	    
     NGC4748 &  12 52 12.4 &  $-$13 24 53& 0.014	& 58.94   & 0.29 &    Sa    &  S1n &  1565  \\	  
     \hline 
     \hline
     \textsc{BLS1 sample} & & & & & & &\\
     \hline
     ESO438G9\footnote{From the bulge disk decomposition, ESO438G9 seems a bulgeless galaxy. This object is therefore rejected from our bulge analysis.}                                                                               
	       &    11 10 48.0  &  $-$28 30 04  &   0.024  & 99.83   & 0.48      &    SBc/d &   S1  &        5000     \\
     F1146     &    08 38 30.8  &  $-$35 59 33  &   0.032  & 131.84  & 0.64      &    Sb    &   S1  &        4300    \\
     IC1816    &    02 31 51.0  &  $-$36 40 19  &   0.017  & 71.31   & 0.35      &    SBa/b &   S1  &         ...     \\
     Mrk0279   &    13 53 03.4  &  +69 18 30    &   0.031  & 123.89  & 0.60      &    Sa    & S1.0  &        6860   \\
     Mrk0290   &    15 35 52.3  &  +57 54 09    &   0.029  & 123.89  & 0.60      &    E     & S1.5  &        2550   \\
     Mrk0352   &    00 59 53.3  &  +31 49 37    &   0.015  & 63.08   & 0.31      &    E     & S1.0  &        3800   \\
     Mrk0423   &    11 26 48.5  &  +35 15 03    &   0.032  & 131.84  & 0.64      &    Sb    & S1.8  &        9000   \\
     Mrk0530   &    23 18 56.6  &  +00 14 38    &   0.029  & 119.91  & 0.58      &    Sa    & S1.5  &        6560   \\
     Mrk0595   &    02 41 34.9  &  +07 11 14    &   0.028  & 111.91  & 0.54      &    Sa    & S1.5  &        2360   \\
     Mrk0609   &    03 25 25.3  &  $-$06 08 38  &   0.032  & 139.74  & 0.68      &    Sa/b  & S1.8  &        ...    \\
     Mrk0704   &    09 18 26.0  &  +16 18 19    &   0.029  & 119.91  & 0.58      &    SBa   & S1.2  &        5500    \\
     Mrk0871   &    16 08 36.4  &  +12 19 51    &   0.034  & 139.74  &  0.68     &    Sb    & S1.5  &        3690   \\
     Mrk0885   &    16 29 48.2  &  +67 22 42    &   0.026  & 103.87  &  0.50     &    SBb   & S1.0  &        ...   \\
     Mrk1126   &    23 00 47.8  &  $-$12 55 07  &   0.010  &  46.48  &  0.23     &    Sb    & S1.5  &        ...   \\
     Mrk1400   &    02 20 13.7  &  +08 12 20    &   0.029  & 119.91  &  0.58     &    Sa    & S1.0  &        ...   \\
     NGC5252   &    13 38 15.9  &  +04 32 33    &   0.022  & 95.79   &  0.46     &    S0    & S1.9  &         2500   \\
     NGC5940   &    15 31 18.1  &  +07 27 28    &   0.033  & 139.74  &  0.68     &    SBc   & S1.0  &         5240   \\
     NGC6212   &    16 43 23.1  &  +39 48 23    &   0.030  & 123.89  &  0.60     &    Sb    & S1    &         6050   \\
     IISZ10    &    13 13 05.8  &  $-$11 07 42  &   0.034  & 139.74  &  0.68     &    ?     &  S1.5 &         3760   \\
     \hline                                                                                  
     \end{tabular}                                           
\end{minipage}                                              
\end{table*}   

To perform the bulge-disk decomposition of these 29 galaxies, we use the two-dimensional profile fitting algorithm \galfit\footnote{http://users.obs.carnegiescience.edu/peng/work/galfit/galfit.html} \citep{P+10}. For each galaxy we iteratively fit three components: a Gaussian profile, a S\'ersic profile and an exponential profile (see also \Aref{app:fit-theory}). These components are aimed at modelling respectively the nuclei, the bulges and the disks of our galaxies. While these HST data of  nearby AGN have small pixel scales (see column 4 in \tref{tab:samples}) enabling us to better constrain the central regions of the galaxies, the FoV of WFPC2 ($\sim 35$\arcsec$\times35$\arcsec) is small relative to the full extent of the galaxies.
This makes it hard to constrain the sky level. 
This issue has already been addressed to some extent by the refinement of our sample selection to objects with $z\geq0.010$.
We cover it further in \Aref{app:fit1} when we discuss our treatment of the possible coupling between the background level and the exponential profile. 
Finally, we analyze the robustness of our fit by studying the effect of saturated regions in the images and the dependence of the fits to the PSF. We detail our iterative fit procedure in \Aref{app:fit1}, our treatment of additional structures such as bars, rings and spirals, and the particular attention given to the background level. 
We give also eight examples of our fits in \Fref{fig:fit-example-nls1} and \Fref{fig:fit-example-bls1}.

The relevant results of our fits are given in \Tref{tab:fits}. 
Among them, the reduced chi-square value $\chi_{\nu}^2$ from the fit, first indicator of its quality. Our mean  $\chi_{\nu}^2$ values are $1.1$ for NLS1s and $1.06$ for BLS1s, reflecting the overall acceptability of the fits.

During this process, we have found two objects for which no acceptable bulge disk decomposition could be performed.
Specifically, the morphology of MRK335 is mainly point-like and is better described by a single high S\'ersic index profile, while ESO438G9 seems to be a bulgeless galaxy also consistent with its morphological type.
We have therefore excluded these two objects from the bulge analysis discussed in Section~\ref{sec:bulge_analysis}, which is performed on 9 NLS1 and 18 BLS1 galaxies.

\begin{table*}
\hspace{2cm}
\caption{Results of the bulge disk decomposition for the NLS1s and BLS1s. Columns: (1) object name; (2) fitted components. The different components  p, g, s, d, b  stand respectively for PSF, Gaussian, S\'ersic, disk (exponential) and background (sky). Components are put in  brackets if one or more of their parameters are kept fixed in the fit; (3) the FWHM$_g$ of the Gaussian component in kpc; (4) the bulge S\'ersic index; (5) $R_b$, the effective radius of the bulge in kpc; (6) $R_d$, the scale radius of the disk in kpc;  (7)-(8) axis ratio of the bulge and the disk components; (9)-(10) B/D and B/T, the bulge-to-disk and bulge-to-total luminosity ratios; (11) $\chi^2_{\nu}$, the reduced $\chi^2$ of the obtained fit. See also \Aref{app:fit-theory}.}
\begin{minipage}{140mm}
     \centering
     \label{tab:fits}
     \begin{tabular}{@{}llrlrrrrrrr@{}}
     \hline
     Object   Name    &    Comp.    &  FWHM$_g$   & $n_b$ & $R_b$ & $R_d$  &$q_b$ & $q_d$ & $B/D$ & $B/T$  & $\chi^2_{\nu}$\\
	              &        	  & (kpc) 	&       & (kpc) & (kpc)  &  &   &     &        \\
     \hline
     \hline
     \textsc{NLS1 sample} &&&&&&&&&& \\
     \hline
     KUG1136  & g+s+d+b  &  0.07  & 1.21 &  0.95   & 3.15  &  0.76   &    0.62   &  0.20 &	0.17  &   1.16 \\       
     MRK0042  & g+s+d+b  &  0.07  & 1.27 &  0.44   & 3.12  &  0.82   &    0.58   &  0.21 &	0.18  &   0.95 \\       
     MRK0335\footnote{From the bulge disk decomposition, MRK335 seems better described by a unique high S\'ersic profile. This object is therefore rejected from our bulge analysis.}
	      & g+s+b    &  0.18  & 3.41 &  1.27   &  --   &  0.93   &    --     &   --  &	 --   &   1.20 \\       
     MRK0359  & g+s+d+b  &  0.08  & 1.41 &  1.19   & 7.45  &  0.67   &    0.83   &  0.12 &	0.11  &   1.03 \\      
     MRK0382  & g+s+d+b  &  0.12  & 1.36 &  0.52   & 3.02  &  0.94   &    0.54   &  0.46 &	0.31  &   1.49 \\       
     MRK0493  & g+s+d+b  &  0.08  & 0.74 &  0.34   & 3.74  &  0.97   &    0.41   &  0.17 &	0.15  &   1.02 \\           
     MRK0766  & g+s+d+b  &  0.06  & 1.88 &  0.15   & 1.21  &  0.74   &    0.44   &  0.13 &	0.11  &   1.05 \\       
     MRK0896  & g+s+d+b  &  0.10  & 2.06 &  0.37   & 2.82  &  0.77   &    0.71   &  0.18 &	0.15  &   1.34 \\      
     MRK1044  & g+s+d+b  &  0.07  & 1.45 &  0.20   & 1.18  &  0.91   &    0.76   &  0.44 &	0.30  &   1.05 \\      
     NGC4748  & g+s+d+b  &  0.07  & 1.93 &  0.25   & 1.89  &  0.98   &    0.68   &  0.25 &	0.20  &   0.68 \\     
     \hline
     \hline
     \textsc{BLS1 sample} &&&&&&&&&& \\
     \hline
     ESO438G9\footnote{From the bulge disk decomposition, ESO438G9 seems a bulgeless galaxy. This object is therefore rejected from our bulge analysis.}
	      &   g+d+b       &   0.06    &     --    &    --      &   1.76      &   --       &  0.52      &    --      &    --     &   1.39 \\
     F1146    &   g+s+d+b     &   0.07    &     3.74  &   0.90     &   2.09      &    0.47    &  0.65      &   1.22     &   0.55    &   1.54   \\ 
     IC1816   &   g+s+[d]+b   &   0.07    &     1.98  &   0.41     &   [4.47]    &    0.92    &  0.67      &   0.11     &   0.10    &    0.84  \\ 
     Mrk0279  &   p+s+d+[b]   &   --      &     2.18  &   2.48     &  11.11      &    0.58    &  0.56      &   0.57     &   0.36    &    0.97  \\
     Mrk0290  &   g+s+d+b     &   0.10    &     4.06  &   0.47     &   2.22      &    0.86    &  0.88      &   0.90     &   0.47    &   1.11   \\
     Mrk0352  &   g+s+d+b     &   0.05    &     4.49  &   0.90     &   1.59      &    0.98    &  0.76      &   0.79     &   0.44    &   0.84   \\
     Mrk0423  &    [s+d]+b    &   --      &     2.13  &   0.43     &   1.72      &    0.69    &  0.68      &   0.73     &   0.42    &   1.62   \\
     Mrk0530  &   [g]+s+d+b   &   0.17    &      2.4  &   1.04     &   4.04      &    0.85    &  0.65      &   0.60     &   0.38    &   0.75   \\
     Mrk0595  &   g+s+[d]+b   &   0.06    &     3.47  &   0.74     &   1.86      &    0.66    &  [0.75]    &   1.92     &   0.66    &   1.30   \\
     Mrk0609  &   g+s+[d]+[b] &   0.15    &     2.28  &   1.48     &   [2.53]    &    0.79    &  [0.95]    &   1.60     &   0.62    &   2.52   \\
     Mrk0704  &   g+s+d+b     &   0.16    &     2.88  &   1.19     &   6.29      &    0.70    &  0.50      &   0.75     &   0.43    &   1.25   \\
     Mrk0871  &   g+s+d+b     &   0.12    &     1.28  &   0.61     &   3.78      &    0.52    &  0.38      &   0.13     &   0.11    &    0.91  \\
     Mrk0885  &   g+s+d+b     &   0.07    &     2.62  &   2.70     &   9.05      &    0.74    &  0.52      &   0.32     &   0.24    &   0.83   \\
     Mrk1126  &   g+s+d+b     &   0.04    &     1.86  &   0.27     &   1.78      &    0.86    &  0.66      &   0.17     &   0.15    &   0.47   \\
     Mrk1400  &   g+s+d+b     &   0.12    &      1.7  &   0.55     &   2.27      &    0.61    &  0.31      &   0.32     &   0.24    &   0.98   \\
     NGC5252  &   g+s+d+b     &   0.07    &     3.9   &   2.42     &   3.20      &    0.53    &  0.44      &   0.58     &   0.37    &   0.35   \\              
     NGC5940  &   g+s+d+b     &   0.10    &     1.23  &   0.30     &   3.77      &    0.89    &  0.66      &   0.06     &   0.06    &   0.99   \\
     NGC6212  &   g+s+d+b     &   0.09    &     1.52  &   1.23     &   2.88      &    0.75    &  0.70      &   0.77     &   0.43    &   0.87   \\
     IISZ10   &   g+s+d+b     &   0.18    &     1.92  &   1.56     &   4.29      &    0.91    &  0.72      &   0.70     &   0.41    &   0.57   \\
     \hline	          
     \end{tabular}
\end{minipage}
\end{table*}

\subsection{Structural properties of NLS1 and BLS1 host galaxies} \label{sec:bulge_analysis}

\subsubsection{The S\'ersic index $n_b$ in the bulges}

\begin{figure}
     \centering
     \includegraphics[width=8cm]{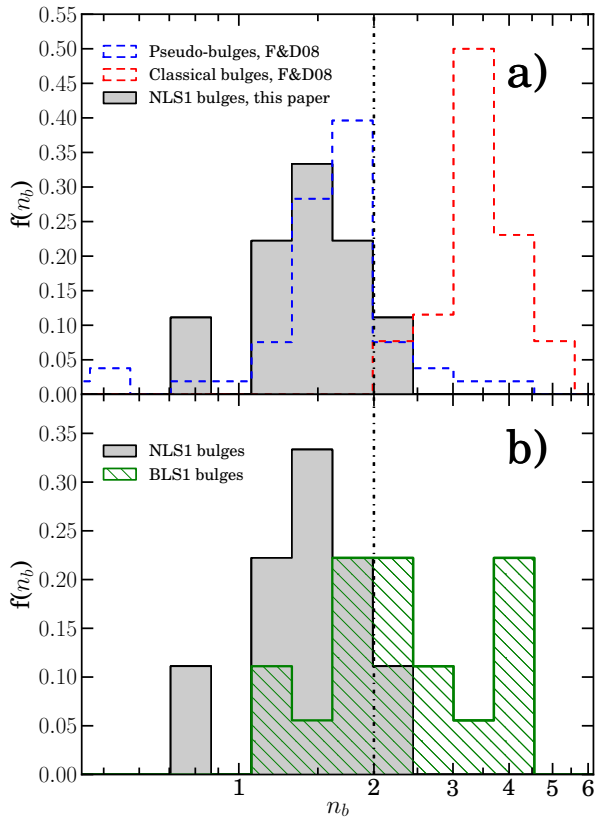}
     \caption{Histogram of bulge S\'ersic indices, $n_b$. a) NLS1 host bulges (9 objects) from our sample compared to pseudo-bulges (53) and classical bulges (26, we do not include their sample of elliptical galaxies) from \citet{FD08}. b) NLS1 host bulges compared to BLS1s host bulges (18 objects).}
     \label{fig:NLS1-FD08}
\end{figure}

In \Fref{fig:NLS1-FD08}, we compare the results obtained for the 9 NLS1 galaxies to the distribution found by \citet{FD08} for pseudo- and classical bulges. According to their results, the mean S\'ersic index of pseudo-bulges is  $1.69$ with only $\sim$10\% of them having an index greater than $2$. For our NLS1 sample, we find  $<n_b> \sim $ 1.48 (and a standard deviation for the distribution of $\sigma_n \sim 0.39$) with none of them significantly exceeding a S\'ersic index of 2.

The bottom panel of \Fref{fig:NLS1-FD08} compares the distribution of NLS1 and BLS1 host bulge S\'ersic indices from the analysis of our samples. 
While the two distributions are clearly different, BLS1 host bulges also tend to have lower S\'ersic indices than the  \citet{FD08} classical bulges. We find $<n_b> \sim $ 2.54, $\sigma_n \sim 0.97$ for BLS1 host bulges, in contrast to $<n_b> \sim $ 3.49 obtained by \citeauthor{FD08} for classical bulges. 
Thus, our results suggest that BLS1s do not have ``pure'' classical bulges, but rather mixed bulges composed of pseudo- and classical components.

It is appropriate to mention here the work of \citet{Lau+07}. They found that, for inactive galaxies, the mean bulge S\'ersic index is $\sim$2.5 or less across the Hubble sequence. These results can also be interpreted as the existence of a large range of composite bulges between the two extreme ``pseudo-'' and ``classical'' bulge types. Putting this result into perspective with our bulge/disk decomposition suggests that the bulges of NLS1 hosts are likely to be  ``pure'' pseudo-bulges, while the bulges of BLS1 hosts appear to be composite bulges and hence have S\'ersic indices distributed around $n_b\sim 2.5$.

\begin{figure}
     \centering
     \includegraphics[width=8cm]{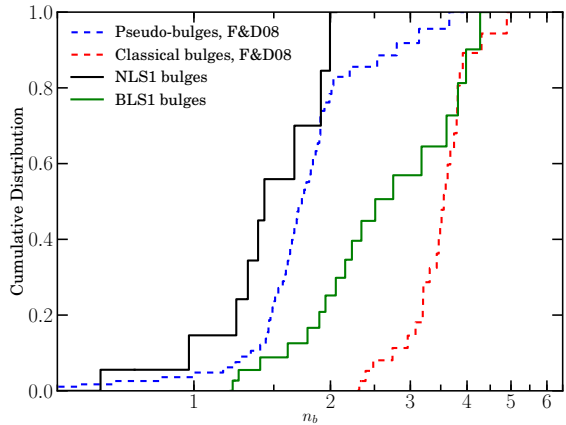}
     \caption{Cumulative distribution vs. the S\'ersic index for pseudo- and classical bulges \citep{FD08}, and NLS1 and BLS1 host bulges, this paper.}
     \label{fig:KS}
\end{figure}

In order to test the significance of the difference between the S\'ersic index distributions of NLS1s and BLS1s, we use the Kolmogorov-Smirnov test.
As such, the cumulative distributions of the S\'ersic index are presented in \Fref{fig:KS}.
The slopes of these distributions reflect the difference in the dispersions given above ($\sigma_n \sim 0.39$ and $0.97$ for the NLS1 and BLS1 samples respectively).
They emphasize that the population of NLS1 bulges appears to be like the population of pseudo-bulges, while the properties of the BLS1 bulges are more widely distributed between pseudo- and classical bulges.
While we cannot fully reject the null hypothesis of the Kolmogorov-Smirnov test, it 
yields a  probability of $<$ 1.2\% that the NLS1s and the BLS1s are drawn from the same parent distribution.
This is a remarkable result that again underlines the connection between black hole and bulge properties.

\begin{figure}
     \includegraphics[height=3.5cm]{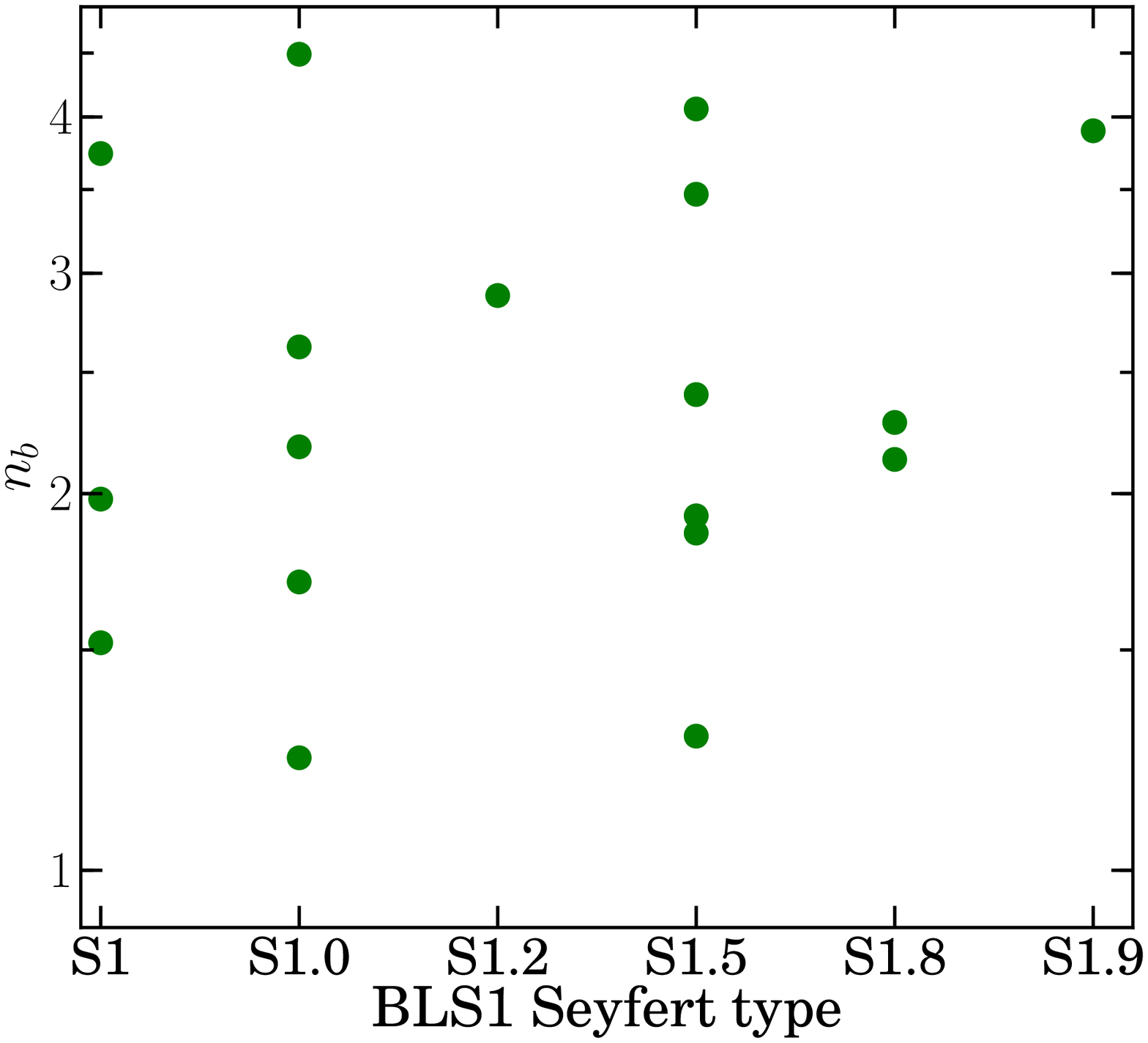}
     \hspace{0.1cm}
     \includegraphics[height=3.5cm]{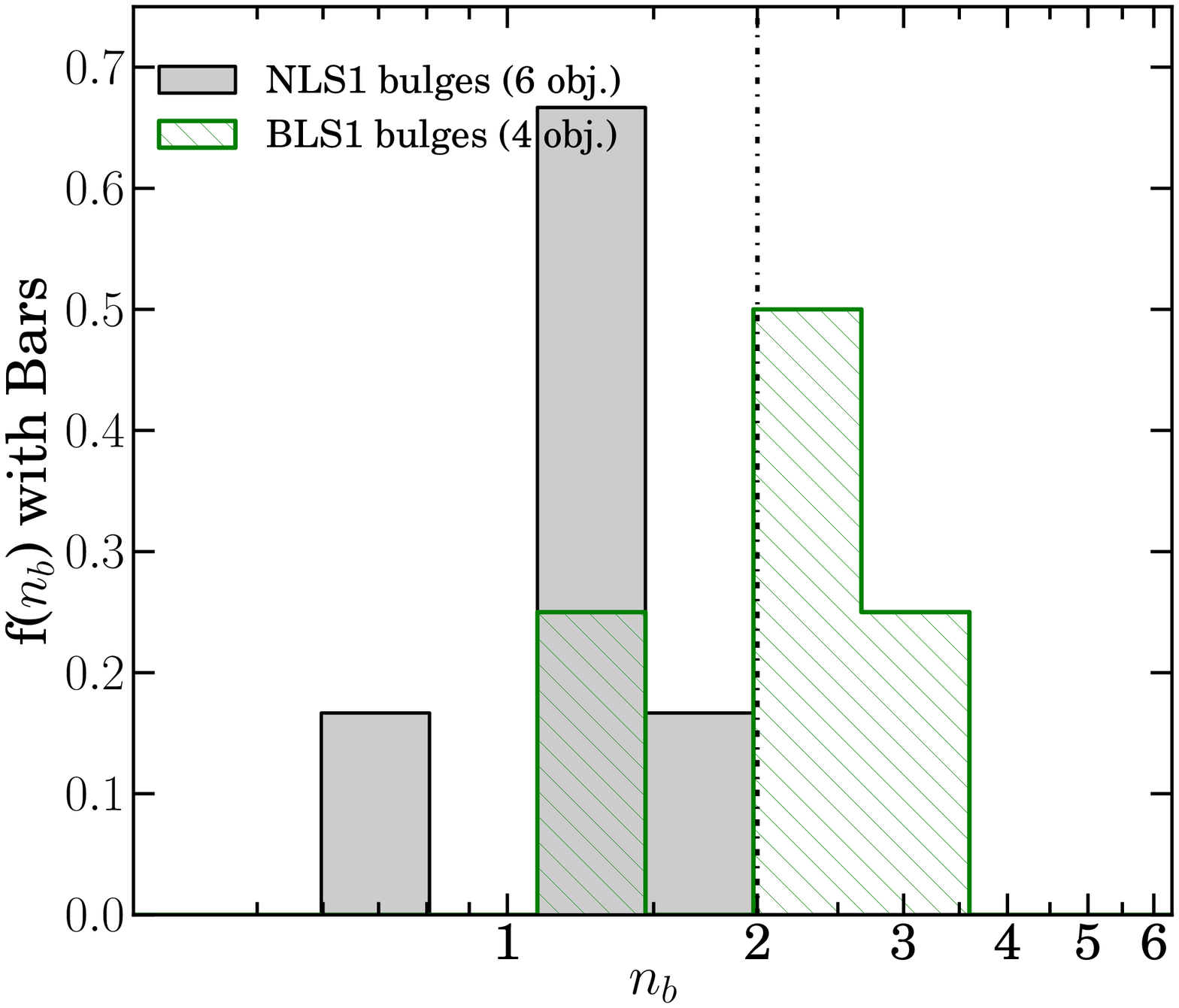}
     \caption{{\it (Left)} BLS1 Seyfert type versus their respective S\'ersic index. Since there is no trend, the S\'ersic index does not seem affected by the Seyfert classification of our BLS1s. {\it (Right)} Histogram of bulge S\'ersic indices, $n_b$, of NLS1 (6 objects) and BLS1 (4 objects) barred galaxies, see also \tref{tab:samples} for the classification and \tref{tab:fits} for $n_b$.}
     \label{fig:S1-n}
\end{figure}

Since a result from \citet{W+09} suggests a possible link between bars and bulges, we verify that the relative bar fractions in our samples have a minimum impact on our result of \Fref{fig:NLS1-FD08}.
Indeed, as described previously in \sref{sec:selection},  our selection of NLS1s and BLS1s  respects the relative  bar fraction observed, \ie $\sim 25\%$ in BLS1 and $\sim 75\%$ in NLS1 galaxies.
Therefore, we consider in \Fref{fig:S1-n} ({\it right}) the distribution of bulge S\'ersic indices of NLS1 and BLS1 galaxies with bars only. While we observe a general small shift towards lower S\'ersic indices (as expected from, e.g., \citealp{W+09}), NLS1 and BLS1 are clearly distinct. Following this observation, we reasonably conclude that although the presence of a bar can be linked to the bulge S\'ersic index, it does not imply that the bulge is a pseudo-bulge.
Thus, our result that NLS1s tend to possess pure pseudo-bulges in contrast to BLS1s, is not a consequence of different bar fractions in the two populations.

Finally, we consider the influence of the Seyfert type of the BLS1s on our S\'ersic measurements. 
Indeed, changes from type~1 to intermediate type Seyferts can be attributed to changes in the ionizing radiation of the AGN \citep[e.g.][]{G90}, or to variation in the absorbing material \citep[][]{MR95,F99}, or as well to different inclination of the host galaxies \citep[e.g.][]{MR95,R+09}.
These effects can explain the large Balmer decrements in the optical spectra of intermediate type galaxies but can also indirectely bias our S\'ersic index measurements. Indeed, in the case of intermediate type Seyferts, a fainter AGN would ease the fit, but an excess of dust in the host galaxy and projection effects (nearly edge-on host spirals) can make fitting the host harder. For these reasons, we check whether there is any systematic effect on our fits by looking at the S\'ersic index versus the Seyfert type of the BLS1s (given in \Tref{tab:samples}), the result is given \Fref{fig:S1-n} {\it (left)}. Since no trend of the S\'ersic index with the BLS1 sub-class can be observed, we can be confident that our fits are not biased by systematic effects related to the intermediate BLS1 classifications.

Similarly, NLS1s are a sub-class of type 1 AGN, for which the Seyfert sub-classification is based solely on the properties of the AGN itself.
But our result also shows that the bulge properties of the NLS1 host galaxies do represent a distinct sub-class of the bulges of BLS1 hosts.

\subsubsection{Bulge prominence and fundamental plane projections}

Building on our fit results, we study the bulge prominence by looking at the size distribution of the bulges, as well as the bulge-to-disk (B/D) and bulge-to-total (B/T) light ratios.
The first objective here is to further explore the differences between NLS1 and BLS1 host bulges. But doing so also enables us to confirm the validity of our bulge-disk decomposition.

\begin{figure*}
     \centering
     \includegraphics[height=4.95cm]{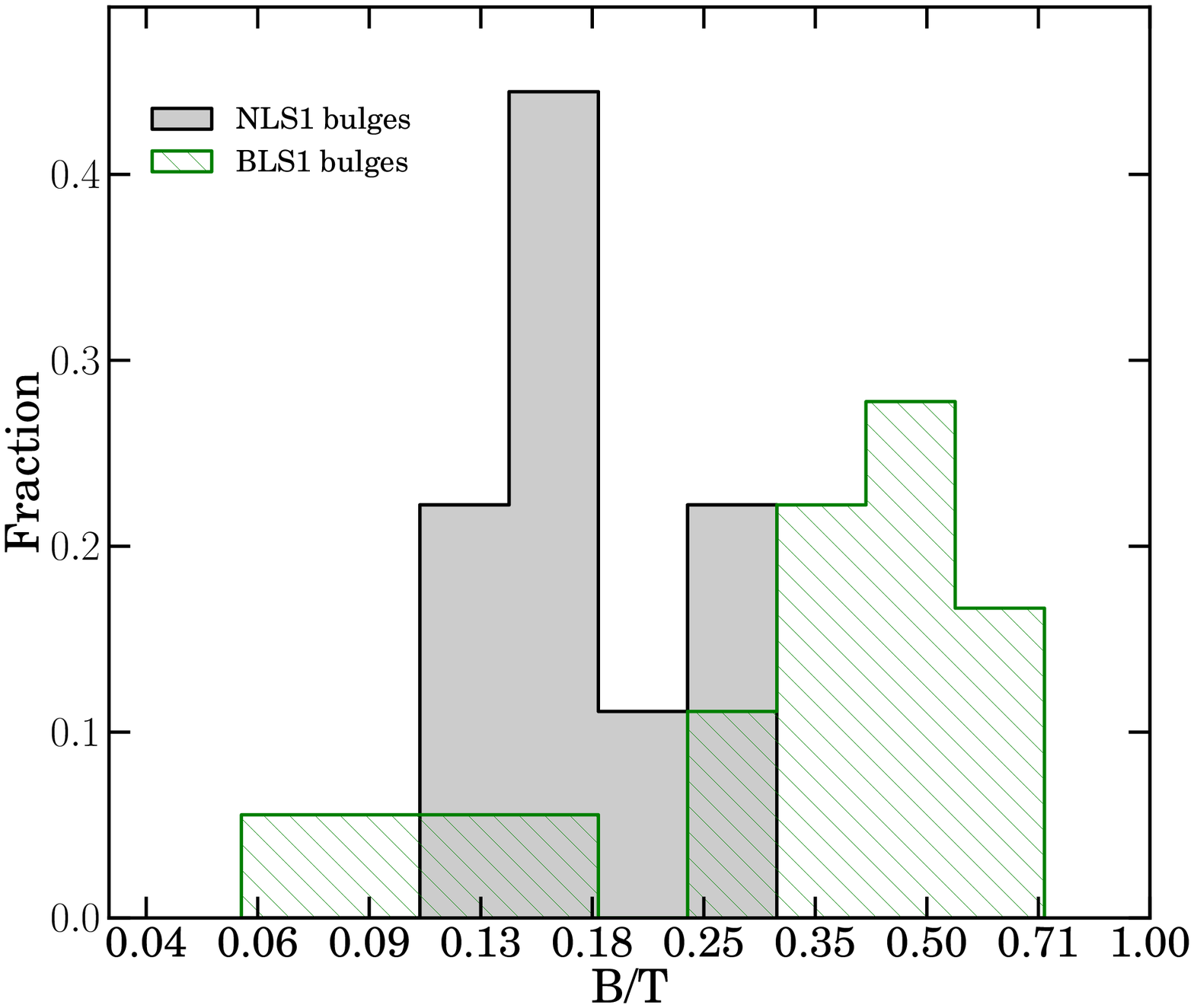}
     \hspace{0cm}
     \includegraphics[height=4.95cm]{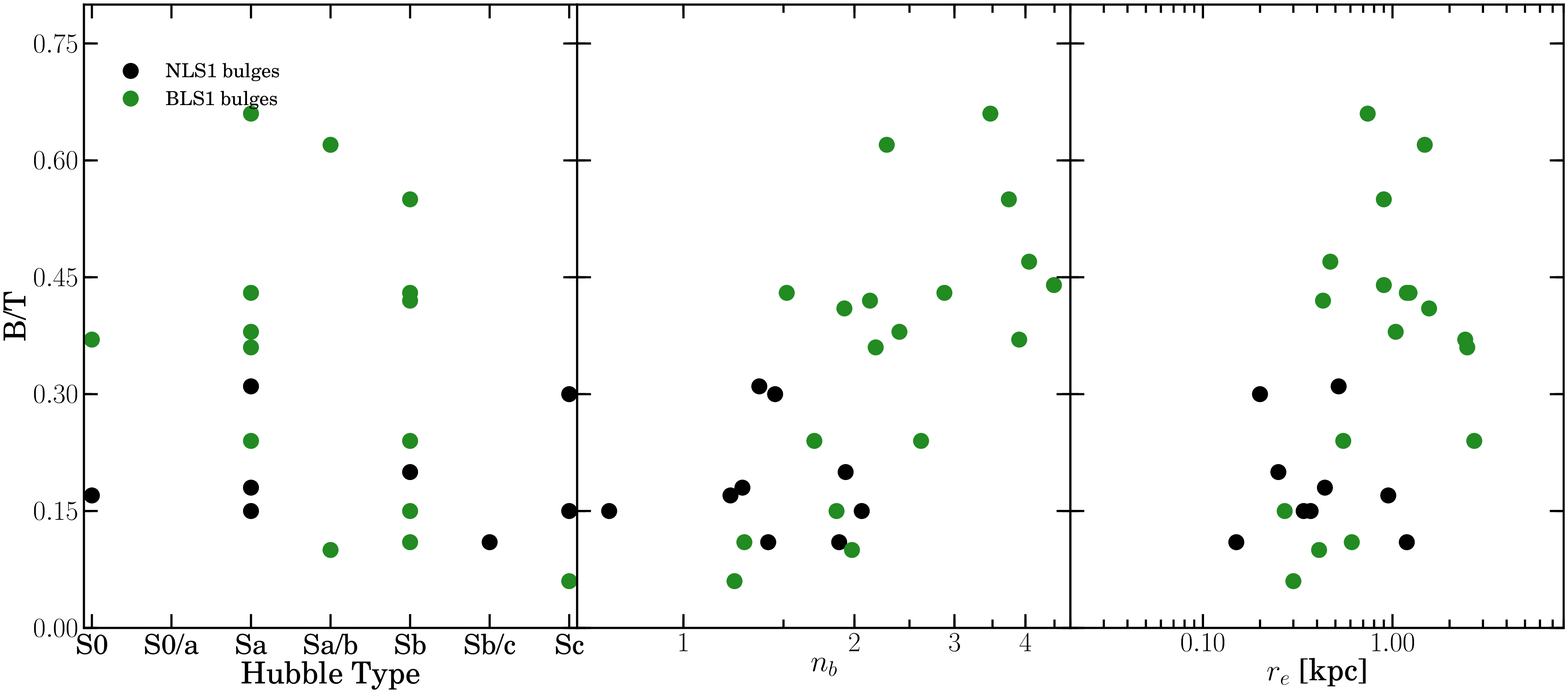}
     \caption{{\it (Left)} Distribution of bulge-to-total light ratio in NLS1 and BLS1s galaxies. According to \citep[e.g.][]{M+10}, NLS1 galaxies would therefore be later type galaxies than BLS1. {\it (Right)} Verification of the mean B/T decrease with the Hubble Type, B/T against the bulge S\'ersic index and against the bulge effective radius. The two last plots confirm the link between B/T light ratios and the prominence of the bulge.}
     \label{fig:BT}
\end{figure*}

We compute the B/D and B/T light ratios using our fit parameters (see \Eref{eq:BD} and \Eref{eq:BT}) given in \Tref{tab:fits}, and present their distribution in \Fref{fig:BT} {\it (left)}.
The median B/T of 0.39 in BLS1s and 0.17 in NLS1s indicates that NLS1 galaxies have lower B/T light ratios than BLS1s.
We compare these distributions to $<$B/T$> = 0.41$ for an average classical bulge and $<$B/T$>=0.16$ for an average pseudo-bulge given by \citet{FD08}.
\citet{G09} also finds similar values.
The B/T ratio therefore provides strong support for our conclusion that NLS1 bulges are pseudo-bulges, while BLS1 bulges are largely composite or classical. 

In the same \Fref{fig:BT} {\it (right)}, we also plot the B/T light ratios versus the Hubble Type, the S\'ersic index, and the effective radius of the bulges. As expected, the mean B/T ratio tends to decrease with the Hubble Type \citep[e.g.][]{GW08,M+10}, and hence it appears that NLS1 galaxies tend to be of later type than BLS1 galaxies (\Fref{fig:BT} {\it (left)}). 
The two last plots illustrate again that NLS1 bulges have less prominent (i.e. smaller, fainter, less cuspy) bulges than BLS1 bulges. 

\begin{figure}
     \centering
     \includegraphics[width=8cm]{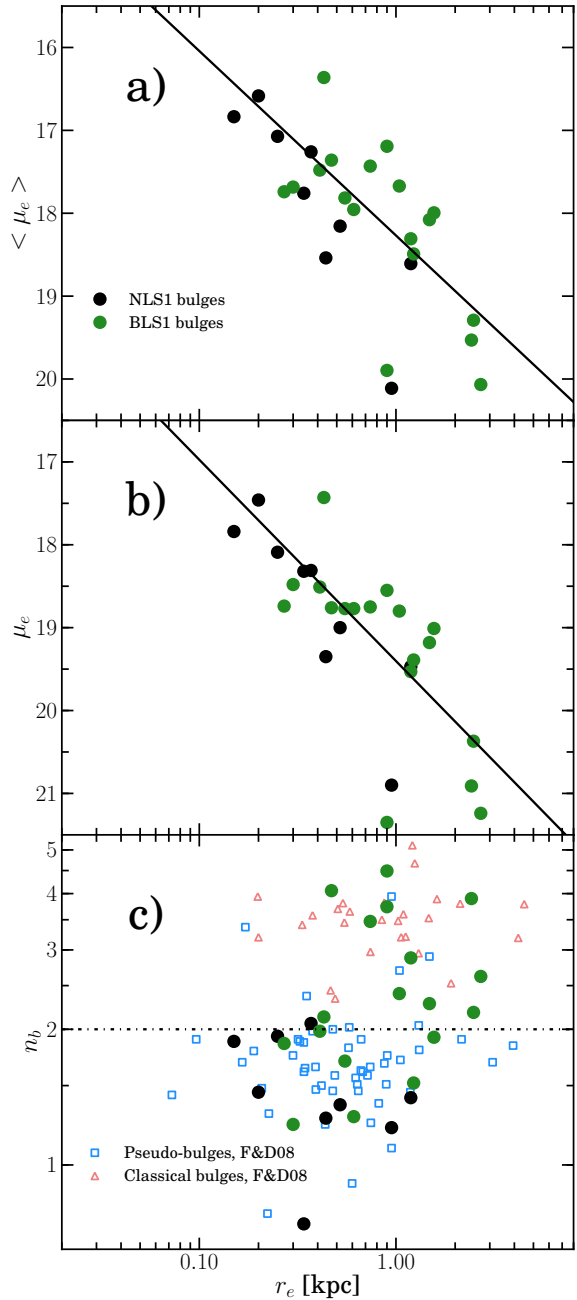}
     \caption{Relations of bulge parameters with the effective radius $r_e$ of the bulges. a) Kormendy relation, \ie the mean surface brightness magnitude within the effective radius versus $r_e$. b) Surface brightness magnitude at the effective radius versus $r_e$. The solid lines are the linear fit found for BLS1s from our sample. The magnitudes are  given in the STMAG system. c) Effective radius vs. S\'ersic index comparing classical and pseudo-bulges from \citet{FD08} with our samples of NLS1s and our BLS1s. The NLS1s seem to lie at the expected $r_e-n_b$ of typical pseudo-bulges.}
     \label{fig:FP}
\end{figure}

While the S\'ersic index is a convincing tool to distinguish pseudo-bulges from classical bulges \citep{FD08}, \citet{G09} uses  the Kormendy relation ($<\mu_e> - r_e$) to identify pseudo-bulges as fainter bulges than predicted by the fundamental plane of elliptical galaxies. 
In \Fref{fig:FP}, we present the Kormendy relation, the surface brightness magnitude at the effective radius, and the S\'ersic index versus the effective radius of the bulge. For \Fref{fig:FP} a) and b) we have overdrawn linear fits to the data for the BLS1 sample. While these do not reveal any marked offset between the two NLS1 and BLS1 classes, NLS1 bulges are systematically fainter than those of BLS1 (i.e. they tend to lie under the line).
This result is consistent with the common structural properties we are finding for the NLS1 class. Finally, \Fref{fig:FP} c) shows NLS1 and BLS1 host bulges, together with pseudo- and classical bulges \citep[from][]{FD08} in the $n_b-r_e$ plane. Again, it shows clearly that NLS1s lie in the region occupied by pseudo-bulges, while BLS1 are spread over the whole range of pseudo- and classical bulge properties.

\begin{figure}
     \centering
     \includegraphics[width=8cm]{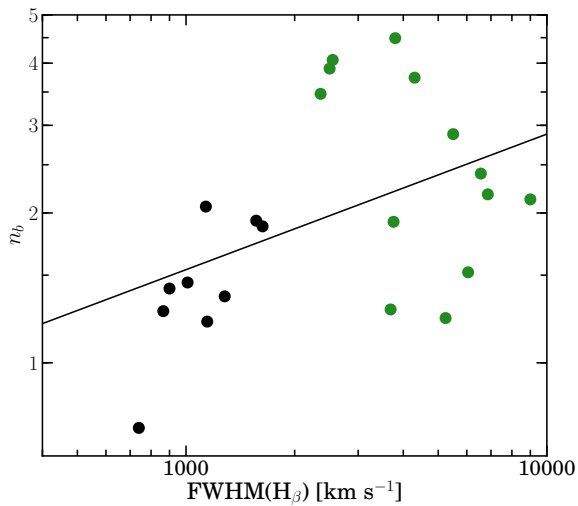}
     \caption{FWHM vs S\'ersic index. NLS1s and BLS1s are represented by black and green dots respectively. The linear fit is based on the two samples. The Spearman's rank correlation coefficient is $\sim0.49$.}
     \label{fig:fwhm}
\end{figure}

Finally,  we also look at the distribution of S\'ersic indices with the FWHM of the broad component of H$\beta$ in \Fref{fig:fwhm}. While the sample is not large enough to make conclusive remarks, we note the existence of a correlation of the FWHM with the S\'ersic index, confirmed by a Spearman's rank correlation coefficient of $\sim0.49$. This correlation  indicates,  at least in the low FWHM($H{\beta}$) range, a possible connection between the bulge concentration and  the broad line region.

\subsubsection{Complementary Studies}

Three other studies support our conclusion about the bulges of NLS1s.

In studying the central engines of NLS1 galaxies, \citet{R+07} perform a bulge disk decomposition of 11 NLS1s galaxies with $z\le0.05$ in the $J$-band and $K_s$-band using adaptive optics data from the 3.6m CFHT. Their mean S\'ersic indices are $<n_J> = 1.52$ and $<n_K> = 1.38$, and  standard deviations $\sigma_n$ of  $\sim 0.44$ and $\sim0.48$ respectively. As the S\'ersic index seems to be, at most, weakly correlated to the photometric band \citep[see e.g.][who compare the S\'ersic index in the V and in the H bands]{FD08}, the \citeauthor{R+07} decomposition supports our results.

In studying low black hole mass systems, \citet{G+08} argue that most of their disk galaxies have pseudo-bulges. Since these systems have small FWHM(H$_{\beta}$), they are also likely to be NLS1s although the relative FeII strengths they found are lower than in classical NLS1s \citep{GH04}.

In a more recent paper, \citet{M+11} study 10 NLS1 galaxies (ACS/HRC using the F625W filter). They perform a similar bulge disk decomposition (\ie they fit a S\'ersic profile for the bulge and an exponential profile for the disk), and also conclude that they have pseudo-bulges. While our conclusions are based mainly on the S\'ersic index, they use the Kormendy relation, as advised by \citet{G09}, to conclude on the pseudo-bulge nature of NLS1 bulges.
Indeed, they do not find systematically low S\'ersic indices but obtain a mean S\'ersic index of $<n>=2.61$ and a large S\'ersic index dispersion $\sigma_{n} = 1.82$. In fact only six out of the ten galaxies in their sample have S\'ersic index values consistent with pseudobulge profiles. 
To try and understand this difference, we note that the \citet{M+11} sample differs from that presented by us here.
Specifically, their sample is found at larger redshift, $<z> \sim 0.24$, leading to a scale in kpc/\arcsec~on average $\sim$4 times larger than in our sample. A large scale limits how well the fit is constrained by the central regions, and confusion in the light distributions between the nucleus and the bulge may arise, possibly leading to higher S\'ersic indices. 

One caveat to these works is the lack of a comparison sample of BLS1 hosts, which our results show is important.
By including one, we show that one should consider the hosts of NLS1s to be a sub-set of all BLS1s, rather than being totally separate.
The distinction between NLS1 and BLS1 hosts is thus that while NLS1 hosts specifically have pseudo-bulges, BLS1 hosts have a range of bulge types including pseudo-bulges, composite bulges, and classical bulges.

\subsection{Secular evolution has always prevailed}

The S\'ersic index distribution and the prominence of the bulge both indicate that, statistically, NLS1 hosts have ``pure'' pseudo-bulges, in contrast to BLS1 galaxies. The consequence of this result \citep{KK04} is that internal secular processes must have dominated the past evolution of NLS1 hosts. And therefore it is from this perspective that one should attempt to explain the particular AGN properties observed in NLS1 galaxies (such as low black hole mass, high accretion rates, etc.).


\section{NLS1 evolution and black hole growth}\label{sec:population}

We explore here the implications of our conclusion that secular processes have dominated the evolution of NLS1 galaxies.
We focus on the issue of the black hole growth of NLS1s, and in understanding whether or not NLS1 galaxies are in a special phase of black hole growth.

\subsection{Expected galaxy populations that have evolved through secular evolution}

Over the last few years, the relative importance of major mergers versus minor mergers and secular processes in driving galaxy formation and evolution has become a key issue in simulations and semi-analytic models.
\citet{Gen+08} and \citet{PEF09} offer two different perspectives to understanding the growth processes of dark matter halos and galaxies.
Both studies conclude that major mergers are not necessarily the main driver of galaxy mass evolution.

\citet{Gen+08} investigate, by analysing cosmological simulations, the growth of dark matter halos. 
They extract halo merger fractions and mass accretion rates from the Millenium simulation in order to study the possible role of major mergers in the evolution of halos from $z\sim2$ to $z=0$.
Following the fate of halos in the mass range 
$11.5 \le \log M_{z=2.2} \le 12.8$, 
they find that $\sim$1/3 of halos which reach z$=$0 have not undergone any major mergers since $z\sim2.2$ and that such halos gain $\gtrsim70$\% of their new mass via mergers less intense than 1:10, demonstrating the importance of non-major merger processes.
In a following paper, \citet{Gen+10} also show that, independently of halo mass, $\sim$ 40\% of the mass in halos have been assembled through smooth accretion.

\citet{PEF09} study galaxy growth by analysing two different galaxy formation models, both also based on the Millenium simulation. Their statistical results are revealing. For both models, they find that only $\leq$49\% of ellipticals, $\leq$3\% of S0s and $\leq$2\% of spirals undergo a  main branch major merger (mass ratio greater than 1:3) in their entire formation history. In other words, $\sim$98\% of spiral galaxies -- which are the most common morphological type of NLS1s host galaxies --  do not undergo any major merger. These results are largely independent of total stellar mass of the galaxy except for ellipticals.
In a further step, they quantify the relative impact of disk instabilities, major mergers and minor mergers on galaxy morphology by determining the stellar mass fraction from each process as a function of the total stellar mass. For both models, they find that instabilities and minor mergers are the main mass contributor, with their relative contributions depending on the treatment of disk instabilities in the models (see  \citealp{PEF09} and references therein).

While it is beyond the scope of this paper to discuss the treatment of the various physical processes in these galaxy formation models, one clear conclusion is that even hierarchical cosmological simulations give rise to large galaxy populations that have evolved through secular processes.
Interestingly, observational studies also reach similar conclusions. 

\citet{W+09} study present day ($D<60$Mpc) spiral galaxies. By performing two dimensional multi-component decompositions of 143 high mass 
($M \ge 10^{10}$\,M$_\odot$) spirals, 
they analyze the bulge S\'ersic index and B/T distributions. Their results highlight the large fraction of bright spirals having B/T$\leq$0.2 ($\sim$69\%) and $n\leq 2$ ($\sim$76\%) where many of them host bars ($\sim$66\%). By comparing their result to theoretical predictions, they find that $\sim66$\% of present day high-mass spirals have not undergone a major merger since $z\leq2$ and likely not even since $z\leq4$. This conclusion conveys the importance of minor mergers (in the present case for a mass ratio $<1:4$) and secular processes since $z\leq4$.

Finally, \citet{Cis+10} recently analyzed the relevance of different triggering mechanisms for AGN activity. Based on visual analysis of 140 AGN and 1264 inactive galaxies with HST imaging, they measure the fraction of distorted morphologies which they take to be a signature of recent mergers. They conclude that the bulk of black hole accretion has been triggered by secular processes and minor interactions since $z\sim1$.

By assessing the role of secular processes, these various theoretical and observational lines of evidence offer a cosmological context to our conclusions concerning NLS1 hosts.
They show that a significant fraction of galaxy mass and a large number of galaxies have evolved from early cosmic times without any mergers.
It is reasonable to suggest that NLS1s, which our analysis shows must have evolved without mergers, may be tracers of this population of galaxies.
Our attempt, in the rest of this section, to put this hypothesis on a more quantitative basis leads us to an estimate of the duty cycle of NLS1s.

\subsection{How common are NLS1s?}\label{ssec:pop}

In the last decade, various surveys of nearby galactic nuclei have found that the fraction of objects classified as AGNs is surprisingly large. They show that, of all local galaxies, approximately 10\% are Seyferts and 40\% can be considered active (see \citealp{Ho08} and references therein).

Looking at the NLS1s, several surveys using optical and X-ray selected samples \citep[e.g][]{W+02,CKG03,Z+06} find that they make up approximately 15\% of Seyfert 1 galaxies. Based on the unified AGN scenario, one can expect that this fraction should apply also to type~2 Seyferts: that $\sim$15\% of Seyfert 2 galaxies may have narrow broad emission lines that are hidden from sight by the obscuring torus. 
Such a possibilty has already been suggested by \cite{ZW06}, who argue that Seyfert~2s without a hidden BLR (i.e. one that cannot be observed in polarised light) are the counter-parts of NLS1s.
Taking this further, one could reasonably argue that the NLS1 definition might extend even to the lower luminosity AGN.

Combining these fractions together leads to the assessment that 2--6\% of local galaxies could be  ``NLS1-like'' galaxies, \ie active galaxies with ``narrow'' broad lines whether obscured or not, the evolution of which has been dominated at all cosmic times by secular processes.

\subsection{Duty cycle of NLS1 black holes}\label{sec:duty}

Our analysis in \sref{sec:bulges} shows that NLS1 galaxies statistically have a bulge S\'ersic index $n_b<2$ and a bulge-to-total light ratio $B/T < 2$. 
The observational study of \citet{W+09}, corroborated by the theoretical study of \citet{PEF09}, argues that approximately 2/3 of local spiral galaxies\footnote{high mass ($M_{\star} \geq 1.0 \times 10^{10} M_{\sun}$)  low-to-moderately inclined ($i<70 \degree$) spirals.} have similar properties to NLS1 hosts. Assuming that each galaxy among this population can potentially undergo nuclear activity and become ``NLS1-like''\footnote{clearly the 2/3 of the local galaxies are not necessarily potential true NLS1s; but, given that their hosts appear to have evolved over cosmic time in a similar way, they might be ``NLS1-like'', as defined in Section~\ref{ssec:pop}, during their accretion phase.}, we can estimate the duty cycle of NLS1s. We argue above that ``NLS1-like'' objects may comprise as much as $\sim$6\% of local galaxies (note that we use the upper end of the range above in order to be conservative later in our estimates of black hole growth). This implies that their duty cycle should be around $\sim$9\%.

Since the Hubble sequence formed at z$\sim$1 \citep{Bergh00, Kaj01, Con+04,Oes+10,Cas+10} with bars becoming numerous at this redshift \citep[][]{A+99,EEH04,Jog+04,MJ07,She+03,She+08}, accretion driven by large-scale bars can only have occurred in the last $\sim$7.7Gyr.
With a duty cycle of $\sim$9\%, this means that NLS1s have actively accreted onto their BHs for $\sim$690Myr.

Assuming their black holes are accreting at the Eddington rate, the e-folding time of their BH build-up is given by the Eddington time-scale, $t_E \approx 4.4\times10^8$ yr.
Therefore, the black hole mass increase is given by \citep[e.g.][]{Vol10b}~:
\be
M_{\mathrm{BH}}(t)=M_{\mathrm{BH}}^{\mathrm{seed}}(t_0)\exp{\left(\dfrac{1-\epsilon}{\epsilon}\dfrac{\tau}{t_{E}} \right)}\mathrm{,}
\ee
where $t$ is the current (observed) time, and $\tau = t-t_0$ is the total accretion duration since the initial time $t_0$.
The radiative efficiency $\epsilon$ is a key parameter, and can have a major impact on the black hole growth rate because it is inside the exponential term.
The standard value of $\epsilon$ is $\sim0.1$.
Adopting this then, and assuming a seed mass $M_{\mathrm{BH}}^{\mathrm{seed}}(t_0)= 10^3 - 10^4 M_{\sun}$, the current mass $M_{\mathrm{BH}} (t)$ could range from $10^9$ to $10^{10}M_{\sun}$.
On the other hand, accretion onto a fast rotating Kerr BH can lead to a 
radiative efficiency $\epsilon \sim 0.2$ or higher \citep{Vol+05,J06}.
For the same $M_{\mathrm{BH}}^{\mathrm{seed}}(t_0)$, this higher radiative efficiency implies a much slower BH growth and leads to current BH masses between $5\times10^5$ and $5\times10^6M_{\sun}$.

Based on these very simple estimates, we reach two conclusions.
Firstly, NLS1s are not necessarily in a special phase of their black hole growth. Their black holes have required 7--8\,Gyr to grow to their current size. To increase their black hole mass by another factor 10 requires -- for a 9\% duty cycle and $\epsilon=0.2$ -- another $\sim$2.8\,Gyr. Thus, despite their high Eddington ratios, NLS1s are not imminently evolving into BLS1s, although they may do this eventually.
Secondly, the low BH masses of NLS1s (with respect to BLS1s) can easily be accounted for by a high radiative efficiency. The theoretical and observational evidence for rapidly spinning black holes in NLS1s is the topic of the \sref{sec:spin}.


\section{Discussion}\label{sec:discussion}

\subsection{NLS1s and highly spinning black holes}\label{sec:spin}
As discussed in the last paragraph, the radiative efficiency of the AGNs have a direct impact on their black hole growth and their final black hole masses. But radiative efficiency is determined by the supermassive black hole (SMBH) spin, which in turn is influenced by the AGN accretion history.
Different scenarios have been studied where SMBH spins evolve through mergers, subsequent  prolonged accretion (constant angular momentum axis of the accreting material, e.g. \citealp{Vol+05,BV08}) or chaotic accretion (random disc orientations with respect to the black hole, e.g. \citealp{KP07,KPH08,BV08}). 
\citet{Vol+05} and \citet{BV08} have shown that if major and minor BH mergers are the sole source of material, then the distribution of BH spins in a $z=0$ galaxy population will reflect that of the initial seed BH spin.
In contrast, they find that prolonged gas accretion triggered by galaxy mergers tends to spin BHs up, and that galaxies where a significant fraction of the BH growth occurs in this mode could have maximal BH spin. However, in such a case, most distant quasars would have high radiative efficiency and would inefficiently grow their black hole. This would require massive BH seeds, in conflict with the Soltan argument.
If instead, the accretion proceeds by short randomly oriented events \citep{KP07,KPH08}, then the spins will tend to be low and lead to high BH masses, resolving the conflict with  the Soltan argument. 
But NLS1 BHs are likely fed via secular rather than merger processes, therefore the angular momentum of the infalling matter could be related to the host structure and hence have a favoured direction. NLS1 SMBHs could be evolving through the prolonged gas accretion scenario. Therefore, NLS1 secular evolution would imply high spins and low BH masses.


From an observational point of view, it appears to be possible to derive BH spins using the Fe K$\alpha$ line in the hard X-ray continuum using accretion disk reflection models. 
While there are several potential reasons that might prevent one from obtaining a good spin constraint for a given source 
(such as too few photons, or too narrow iron line hampering the fit to pick out the role of relativistic contributors, \citealp{Bre07})
several recent works have been able to derive formal constraints on BH spin for a number of Seyfert galaxies \citep[e.g.][]{Bre07,Fab+09,Min+09,Pon+10}. Among these measurements, we find that most of them are for NLS1s, and that the derived spin values are very high. While there are still too few measurements to draw any firm conclusion, these results are already suggestive that NLS1s could have highly spinning BHs owing to the prolonged disk accretion onto their BHs.

Finally, as we have already touched on in \sref{sec:duty}, a high spin leads to a high mass-to-energy conversion or radiative efficiency (because the last stable orbit is closer to the horizon of the BH), and hence to a slow BH growth.
Therefore, the high Eddington ratio typically observed in NLS1s could be a signature not just of highly accreting BHs, but also of rapidly spinning BHs. 
Given that pseudo-bulges are one consequence of secular evolution in a galaxy, and that another consequence is that the SMBH should be rapidly spinning, 
our result that NLS1 hosts have pseudo-bulges leads to the prediction that a large fraction of NLS1 BHs should have very high spin. Future X-ray missions may be able to test this.

\subsection{Black hole - bulge scaling relation}

One question concerning NLS1s that has received much attention is whether they follow the M$_{\mathrm{BH}}-\sigma_{\star}$ relation or are offset under it. 
The  M$_{\mathrm{BH}}-\sigma_{\star}$, or more generally the BH-bulge scaling relations, are often interpreted as physical evidence for the co-evolution of the central BHs with the galactic bulges.
The case of NLS1s is rather controversial. On one hand, many studies suggest they may reside below the relation, in which case they could be evolving onto it \citep[e.g.][]{MKC01,BZ04,GM04,MG05b,MG05a,Z+06}. 
On the other hand, several different studies place them on the relation once contaminating effects have been corrected for, such as [OIII] line\footnote{this line is often used as a surrogate of the stellar dispersion $\sigma_{\star}$, see \citet{Nel00}.} broadening due to outflows (see \citealp{B+04} and \citealp{KX07}), or radiation pressure \citep[as proposed by][]{M+08}.

Current developments regarding the M$_{\mathrm{BH}}-\sigma_{\star}$ relation highlight that it may not be universal (common to all morphological types), and that perhaps one should distinguish between barred and barless galaxies, disks and ellipticals \citep[e.g.][]{G08,Hu08,GL09,G+11}, classical bulges and pseudo-bulges \citep[e.g.][]{N+10,S+10prep,KBC11}. In fact, barred, disk galaxy bulges and pseudo-bulges appear either to lie below the relation or to scatter around it.
Additionally, on a more theoretical side, some authors \citep{Pen07,J+11} suggest that the BH-bulge scaling relations could be non-causal (their origin would not invoke a physical coupling between the SMBH and the galaxy) but rather would be naturally produced by the merger-driven assembly of bulge and BH masses,
 and therefore galaxies with pseudo-bulges would not be expected to obey the same relation \citep{J+11}.

Our result that NLS1s have pseudo-bulges suggests that we should expect these AGN to lie in the same region as inactive galaxies with pseudo-bulges, that is scattered around and below the M$_{\mathrm{BH}}-\sigma_{\star}$ relation. 
It is not yet understood how -- or whether -- black hole and bulge growth are linked when secular processes drive their evolution.
Thus, while it is clear that their black holes are still growing, we cannot predict where they will end up on the M$_{\mathrm{BH}}-\sigma_{\star}$ plane.

\subsection{Evolutionary scenarios}

Several authors have suggested different links between NLS1 galaxies and other AGN types in evolutionary sequence contexts. We briefly discuss them in the light of our results. 

\citet{Mat00} argues that NLS1 galaxies might be in an early stage of evolution owing to their small growing black holes and higher Eddington rates. This proposition is not inconsistent with our results.
Nevertheless, it illustrates a different perspective. Either NLS1s would have their nuclear supermassive black holes recently formed and NLS1s would be young objects evolving into BLS1s \citep{Mat00,M+11}, or NLS1s would not be in any special phase of their evolution but simply  have BHs that are growing slowly due to their duty cycle and spin.
However, both perspectives agree that NLS1s galaxies have pseudo-bulges and that their black hole growth is driven by secular processes as opposed to mergers at high redshift.

\citet{KIN07} proposed an evolutionary track from type 1 ULIRG, to NLS1 to BLS1. The connection between ULIRG and NLS1 appears contradictory with our results. Indeed, local ULIRGs are the result of galaxy mergers, while we have argued that, based on the host properties, NLS1s have a secular driven evolution.

\citet{ZW06} study Seyfert~2s with and without a hidden BLR (i.e. presence or absence of BLR in polarized light) and suggest that non-HBLR Seyfert~2s are the counterparts of NLS1s viewed at high inclination angles. 
In their subsequent paper, \citet{WZ07} propose an evolutionary sequence of the narrow objects to broad line AGN considering time evolution of the black hole mass and the accretion rates. While it is not inconsistent with our results, the distribution of bulge properties in BLS1s suggests that not all BLS1s come from NLS1s that have evolved secularly, but that the BLS1 population should include galaxies that have undergone interactions and mergers.
While  \citet{WZ07} propose a secular evolution from NLS1s to BLS1s (NLS1s would be an early AGN phase and would evolve to BLS1s during the AGN activity time), \citet{ZZT09} proposed a similar scenario, but where NLS1s would be produced by mergers of smaller galaxies compared to BLS1s and could evolve to BLS1s only if they encounter more mergers to grow them. 
This last scenario appears contradictory to the results of the present paper.

While at this point there is no consensus on the cosmic evolution of NLS1s, our results suggest that they are a special case in which the evolution has been dominated at all time by secular processes.
Thus, if the BHs in NLS1s continue to grow, they must eventually become broad-line AGN, and thus become part of the BLS1 population.
However, our results also show very clearly that not all BLS1s have grown in this way, and that mergers have played a role in the evolution of the BLS1 population.
In this respect, perhaps the most enlightening question would be: when  
NLS1s evolve into BLS1s, will they be distinguishable from systems classified as BLS1s but having undergone galaxy interactions and mergers?
Perhaps one can already begin to address this by studying the BLS1s with pure pseudo-bulges, and asking whether they have definable characteristics that differ from the BLS1 population as a whole.

\section{Conclusions}\label{sec:conclusions}

From a review of the literature, we show that secular evolution in NLS1 galaxies is a powerful and on-going process on all scales, in contrast to BLS1 galaxies. To assess the role of secular processes in the past evolution of NLS1 galaxies, we examine their bulge properties by performing bulge-disk decompositions on NLS1 and BLS1 galaxies with archival HST images. The results indicate that NLS1 host bulges are pseudo-bulges and distinct from the much broader population of BLS1 bulges.
From these results, we conclude that NLS1s represent a class of AGN in which the black hole growth is, and has always been, dominated by secular evolution.

Such an evolutionary mode signifies also a different black hole growth mode in NLS1s than in merger-built systems. Interestingly, simulations of prolonged disk-mode gas accretion onto black holes show that the most efficient way to spin-up a black hole is through smooth accretion of material. In this light, our results suggest that NLS1 galaxies should possess highly spinning BHs which is indeed what has so far been observed. 

Our picture of the NLS1 galaxy phenomenon can be expressed as follows. The activity in NLS1 galaxies is, and always has been, powered by internal secular processes.
This has lead to the growth of a pseudo-bulge. 
It is also characterized by a disk-mode accretion onto the central object, which tends to spin up the black hole.
This leads to high radiative efficiency of the accreting material, therefore reducing the actual mass accreted onto the black hole and slowing its growth.
The high radiative efficiency could in part explain the high Eddington ratios and small black hole masses of NLS1s.

This picture can be tested by analyzing the angular momentum in NLS1 bulges to assess definitively their pseudo-bulge nature. And studying the kinematics in the central part of NLS1s would help to understand how gas is transported to their central regions, what the mass inflow rates are, and the role played by angular momentum. In parallel, systematic measurements of black hole spins by future X-ray missions would also shed light on the growth of their black holes at the smallest scales.

\section*{Acknowledgments}
We thank E. Sani and S. Mathur for useful discussions. We are grateful to an anonymous referee, A. Barth, E. Cameron, A. Graham, K. Jahnke, A.King  and B. Simmons  for useful and interesting comments. 
GOX acknowledges support from and participation in the International Max-Planck Research School on Astrophysics at the Ludwig-Maximilians University.


\label{lastpage}
\bsp

\appendix

\section{S\'ersic light profile}\label{app:fit-theory}
The S\'ersic power-law intensity profile is  frequently used  in the study of galaxy morphology. It has the following functional form (e.g. \citealp{C+93,P+10})
\be
I(r)  =  I_e \exp\left[  -\kappa \left(  \left(\dfrac{r}{r_e}\right)^{1/n}  - 1\right)\right]\mathrm{,}
\ee
where $I_e$ is the surface brightness at the effective radius $r_e$. The parameter $r_e$ is known as the effective, or half-light, radius, defined such that half of the total flux lies within $r_e$. The parameter $n$ is the S\'ersic index: with $n=4$ the profile is the de Vaucouleurs profile typical of elliptical galaxies, $n=0.5$ gives a Gaussian, and $n=1$ is the exponential profile typical of disks. Finally, the parameter $\kappa$ is in fact coupled to $n$ and is not a free parameter.

The exponential disk profile is more frequently described by the compact form
\be
I(r)  =  I_0 \exp\left(- \dfrac{r}{r_s} \right)\mathrm{,}
\ee
where $I_0$ is the central surface brightness, $I_0=I_e e^{\kappa}$, and $r_s$ the scale radius given by $r_e=1.678 r_s$, or more generally by $r_e=\kappa^n r_s$.

\subsection{S\'ersic luminosity ratios}
The bulge-to-disk (B/D) and bulge-to-total (B/T) luminosity ratios are relevant quantities for the study of galaxy morphology. These ratios tend to decrease from early- to late- type spirals (e.g. \citet{M+10}). The B/D ratio, where the disk is described by an exponential profile, can be expressed analytically as
\be \label{eq:BD}
\dfrac{B}{D} = \dfrac{n_b \Gamma(2n_b) e^\kappa}{\kappa^{2n_b}} \left( \dfrac{q_b R_b^2}{q_d R_d^2}\right) \left( \dfrac{I_e}{I_0}\right)\mathrm{,}
\ee
and the B/T as
\be \label{eq:BT}
\dfrac{B}{T} = \dfrac{n_b \Gamma(2n_b) e^\kappa / \kappa^{2n_b} q_b R_b^2 I_e } { n_b \Gamma(2n_b) e^\kappa / \kappa^{2n_b} q_b R_b^2 I_e  + q_d R_d^2 I_0  }\mathrm{,}
\ee
where the subscripts $b$ and $d$ refer to the bulge and the disk respectively. $I_e$ is the effective surface brightness of the bulge and $I_0$ is the central surface brightness of the disk, $R_b$ is the effective radius of the bulge and $R_d$ the scale radius of the disk, and $q_b$, $q_d$ are the axis ratios of the respective profiles. 
To calculate the parameter $\kappa$, we use the analytic expansion Equation 18 from \citet{CB99} valid for $n>0.36$.

\section{Fit procedure} \label{app:fit1}
To accomplish the bulge-disk decomposition, we use the two-dimensional profile fitting algorithm \galfit \citep{P+10}, which allows, by minimizing the $\chi^2$ value, to model the light profiles using e.g. S\'ersic profiles.

Since we fit HST images, we generate the PSFs with the \tinytim\footnote{http://www.stsci.edu/software/tinytim/} code for the WFPC2 camera as well as the F606W filter. 
For each image, we create a PSF referenced to the galaxy center with an uniform weight along the wavelength range.

Initial parameters are estimated from visual inspection of the images (positions, P.A., ellipticities, radii, etc.). Nevertheless, our iterative fit procedure, as described hereafter, ensure that the choice of initial estimates do not influence the final results.

\begin{figure}
     \centering
     \includegraphics[width=8.5cm]{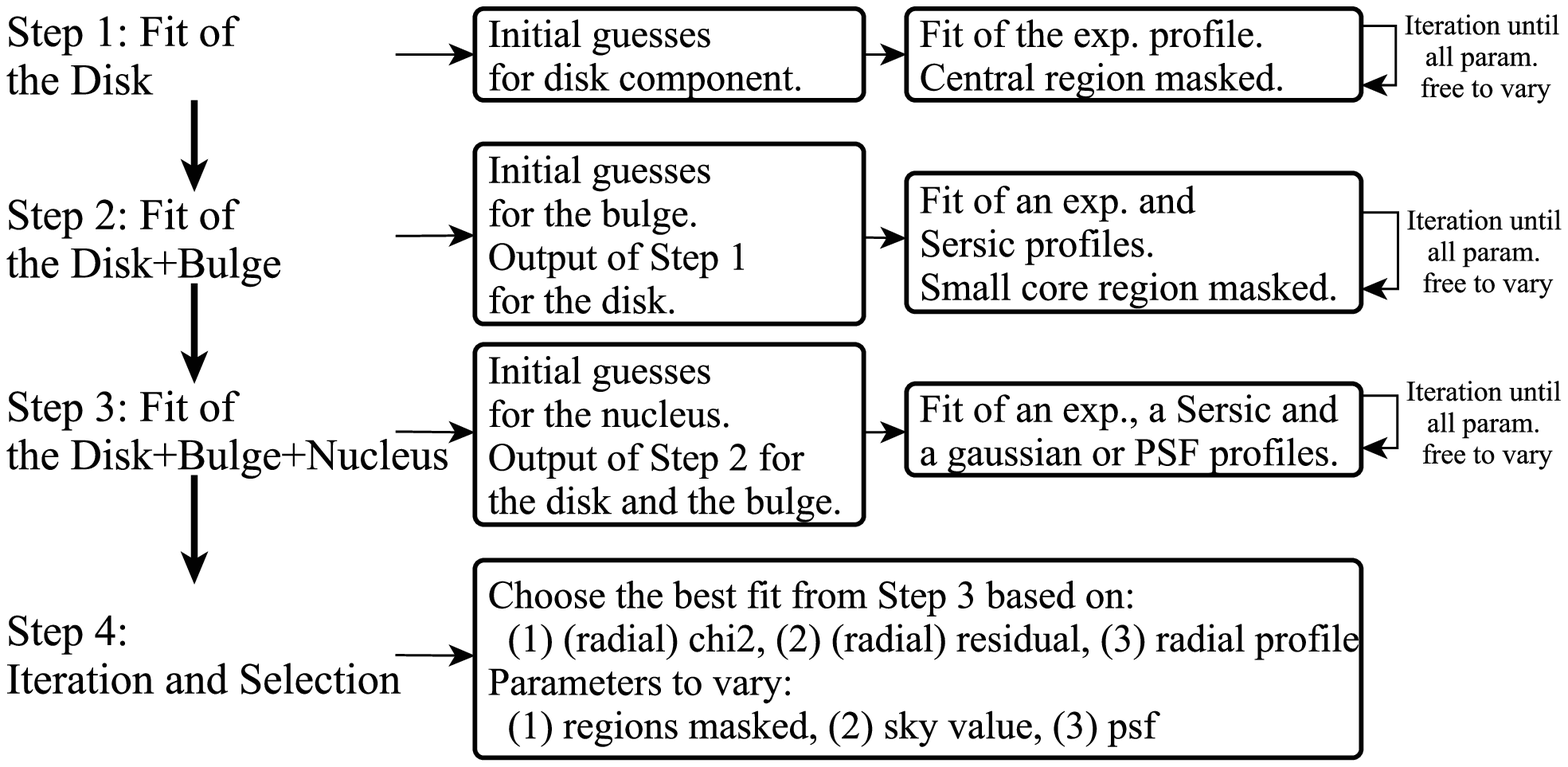}
     \caption{Our iterative fitting procedure.}
     \label{fig:proc}
\end{figure}

Our procedure can be divided in 4 steps as summarized in \fref{fig:proc}. For each step we create appropriated masks to remove regions from the fits. 

1. We start by fitting only the disk of the galaxies using an exponential profile and masking the central region. 

2. Once a reasonable model is obtained, we fit an additional S\'ersic component to model the bulge. Our initial estimate of the  S\'ersic index is $n_b=2$ (\ie the index threshold between pseudo- and classical bulges), while the other parameter initial estimates are based upon visual inspection. When fitting, we first keep  the outputs of our step 1 fixed in the fit and then free to vary (together with the S\'ersic profile parameters). During this step we mask only the core  of the galaxies  (5 to 10 pixels radius) to avoid possible influence of the central nucleus or AGN source. 

3. We then fit an additional Gaussian (with initial FWHM of 2.5 pixels) to model the AGN and iterate, if necessary, until all parameters are free to vary. As a non-negligible fraction of the images  presents a saturated core with charges leakage along the columns (34\% in total, 50\% in the NLS1 and 26\% in the BLS1 samples), we mask carefully these pixels. We also verify the resulting FWHM of the Gaussian and its axis ratio. An additional motivation to use a Gaussian instead of a PSF is  to model any nuclear star cluster which, if not accounted for in the fit, will artificially increase the bulge S\'ersic index.

During these three steps, we judge the quality of the fit based on the residual images, the $\chi^2$ values, and the parameters values (ensuring that they are physically meaningful~: the magnitude of the components have to be greater than the sky level, the S\'ersic index value should be acceptable, in particular $n_b<8$, and the physical radii $R_b$ and $R_d$ should be reasonable). 

4. At the end of the third step, we additionally look at the radial profiles of the model, its components, and the original image. These plots, together with the radial residual and the radial $\chi_{\nu}^2$ enable us to diagnostic possible problems and identify influences of non-fitted structures.
Upon examination of these plots and the image residuals, we decide whether it is necessary to mask relevant structures such as rings or spirals. We refine the masks and the fits until the radial profile of the model does not appear to be influenced by these structures but do translate correctly the bulge and the disk components. 

\begin{figure*}
     \centering
     \subfigure{\includegraphics[height=8.5cm]{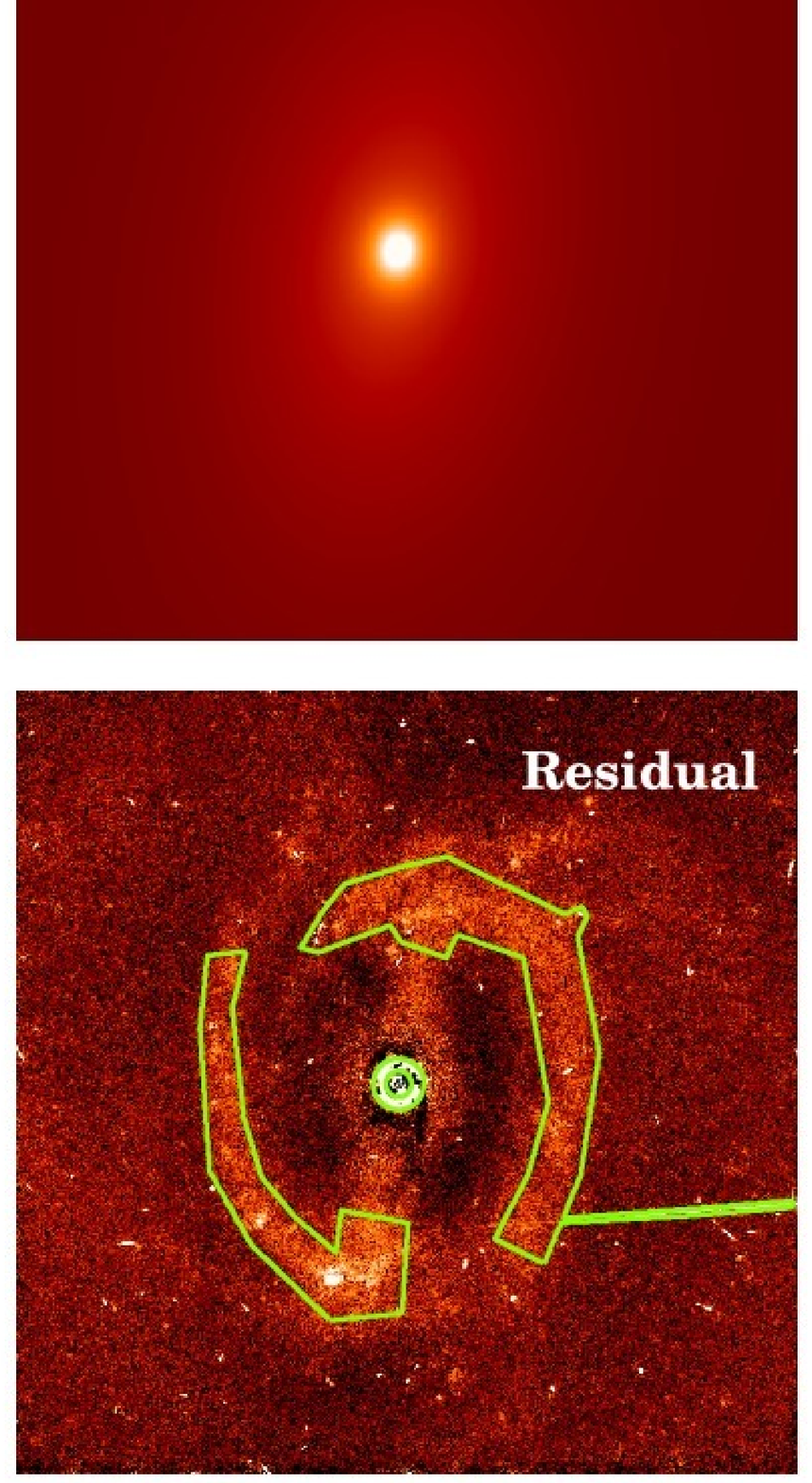}
     \includegraphics[height=8.5cm]{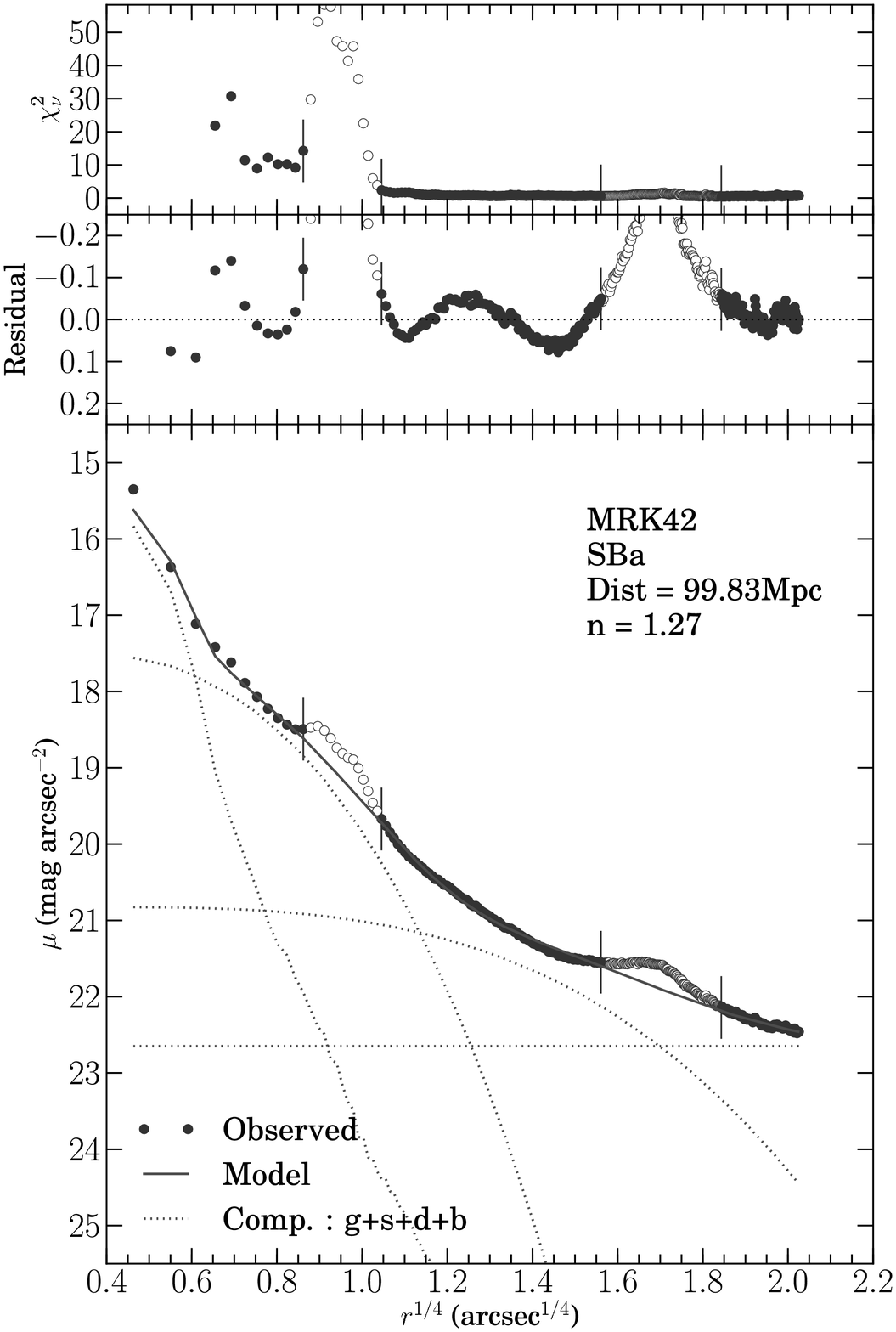}}
     \subfigure{\includegraphics[height=8.5cm]{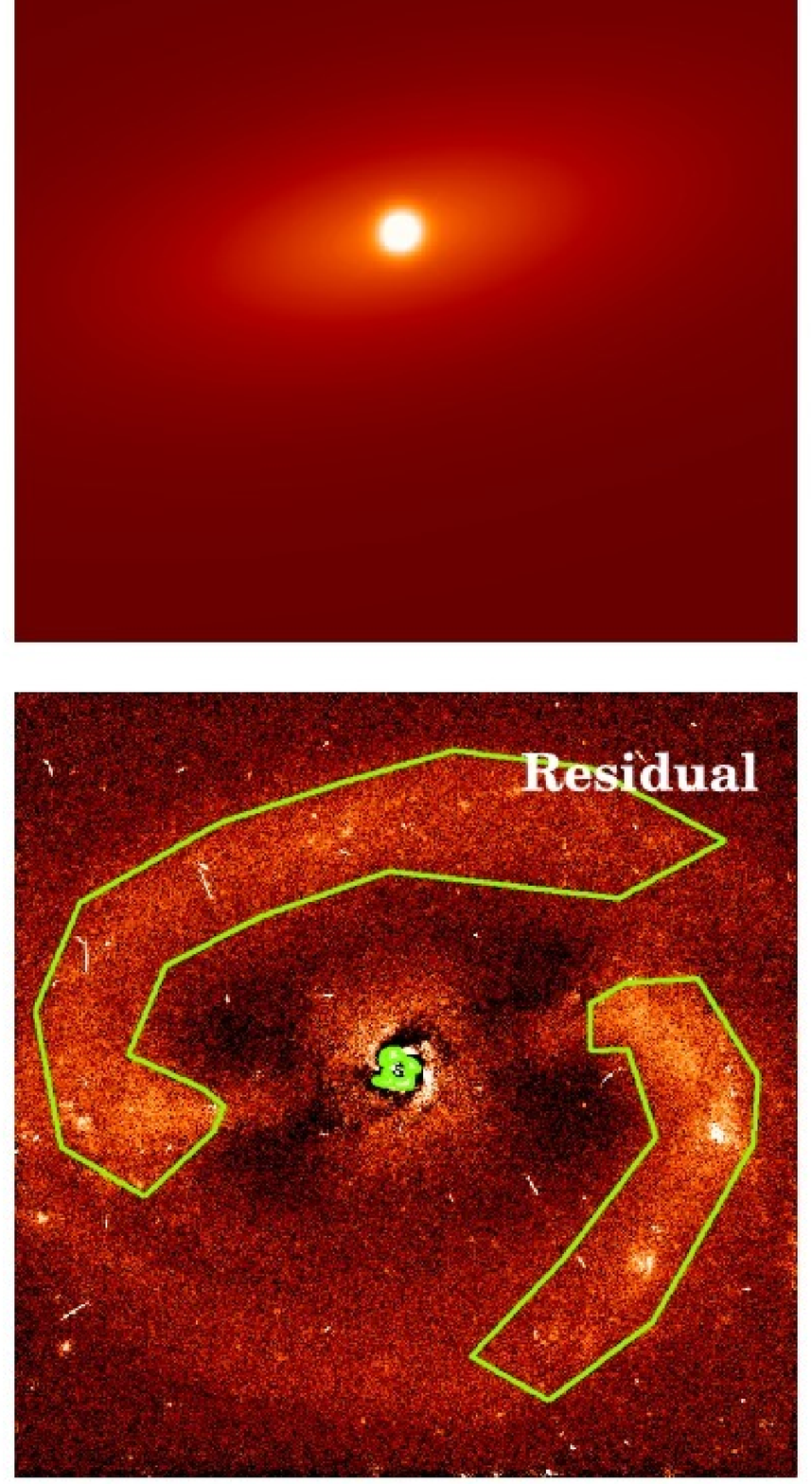}
     \includegraphics[height=8.5cm]{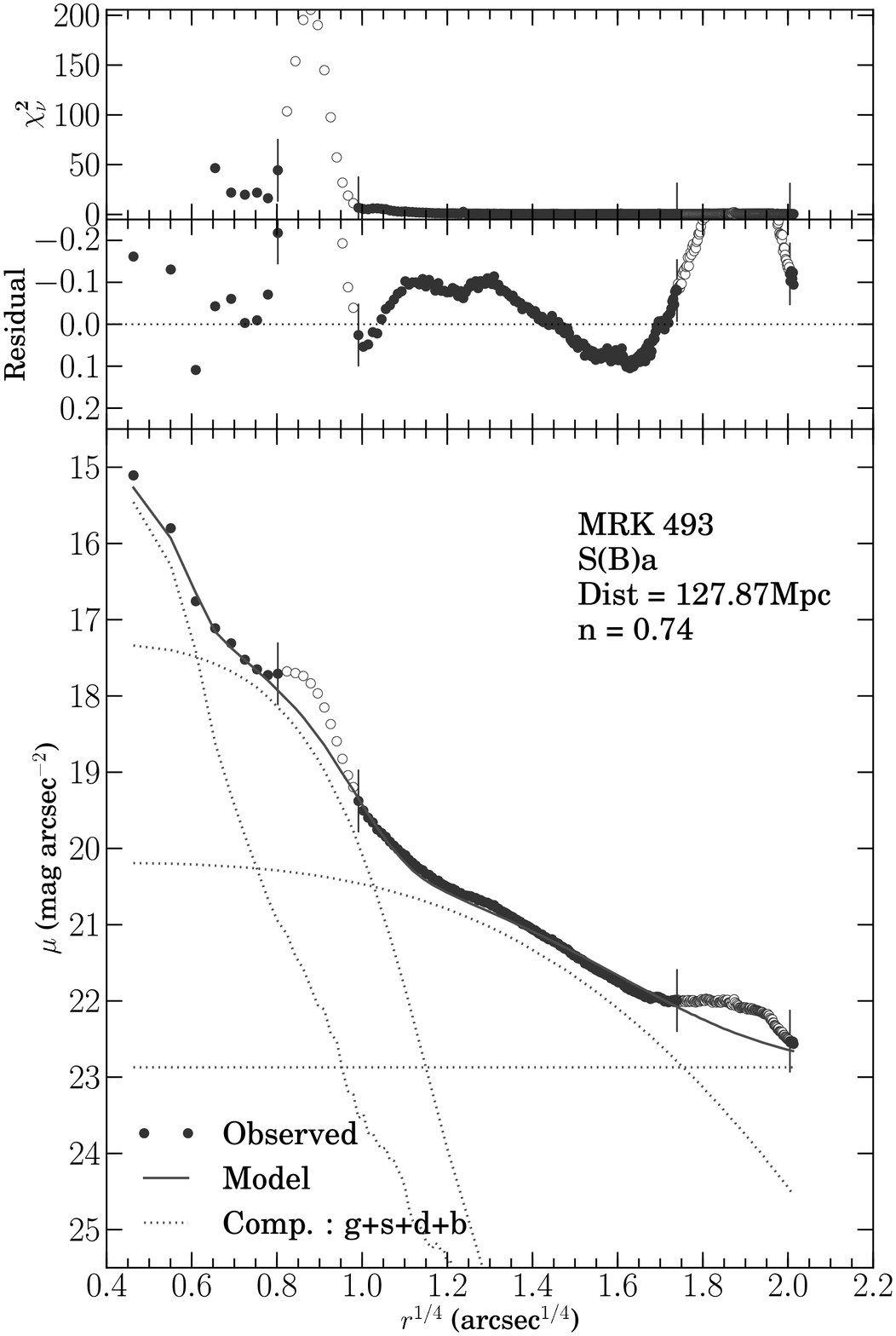}}
     \subfigure{\includegraphics[height=8.5cm]{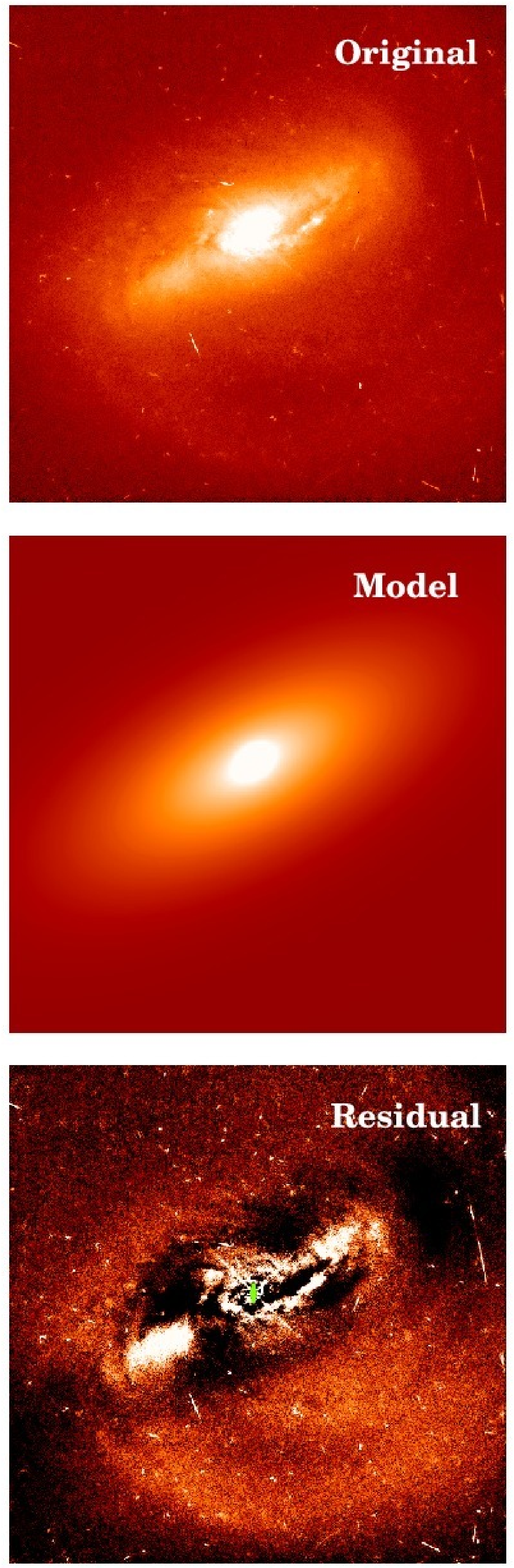}
     \includegraphics[height=8.5cm]{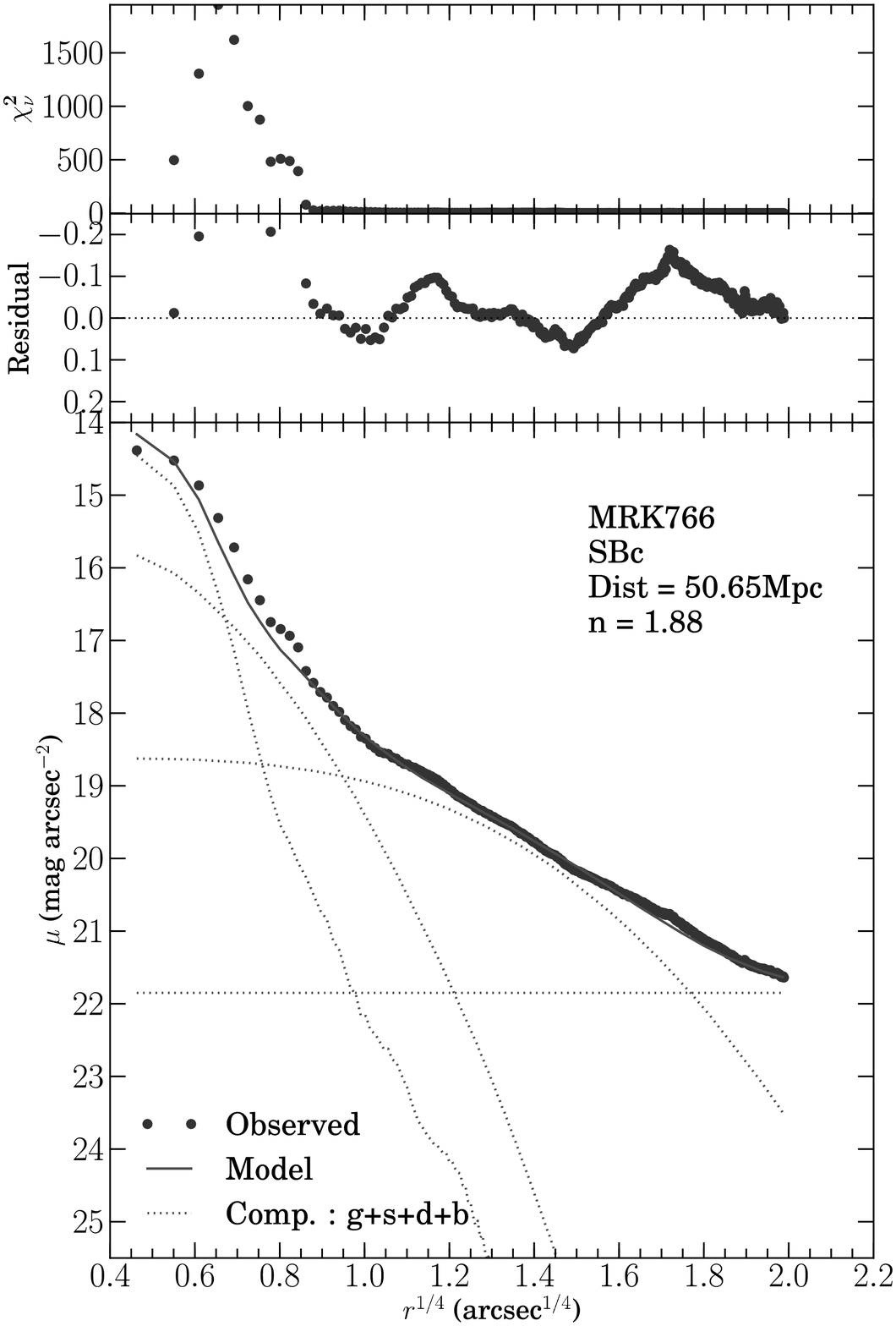}}
     \subfigure{\includegraphics[height=8.5cm]{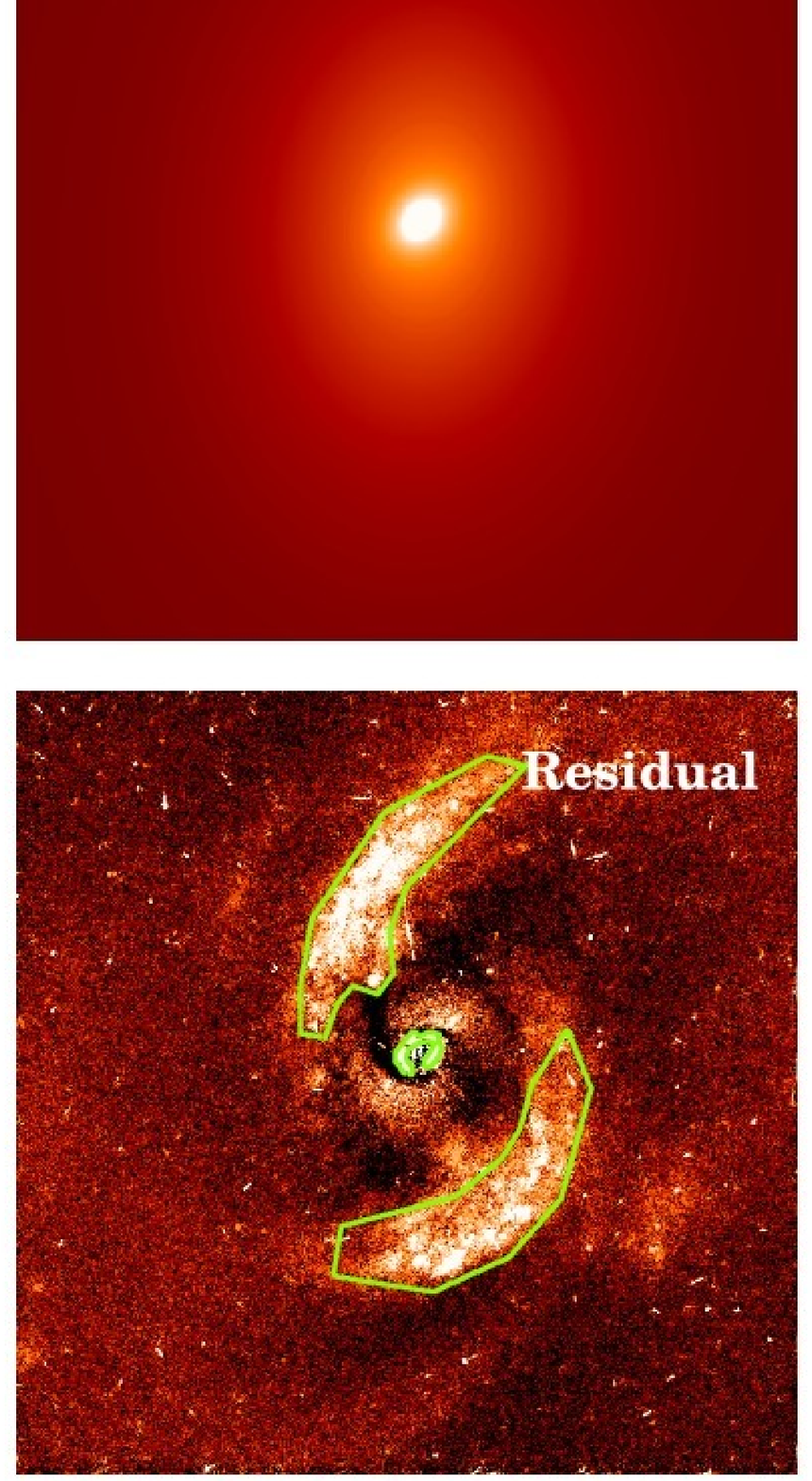}
     \includegraphics[height=8.5cm]{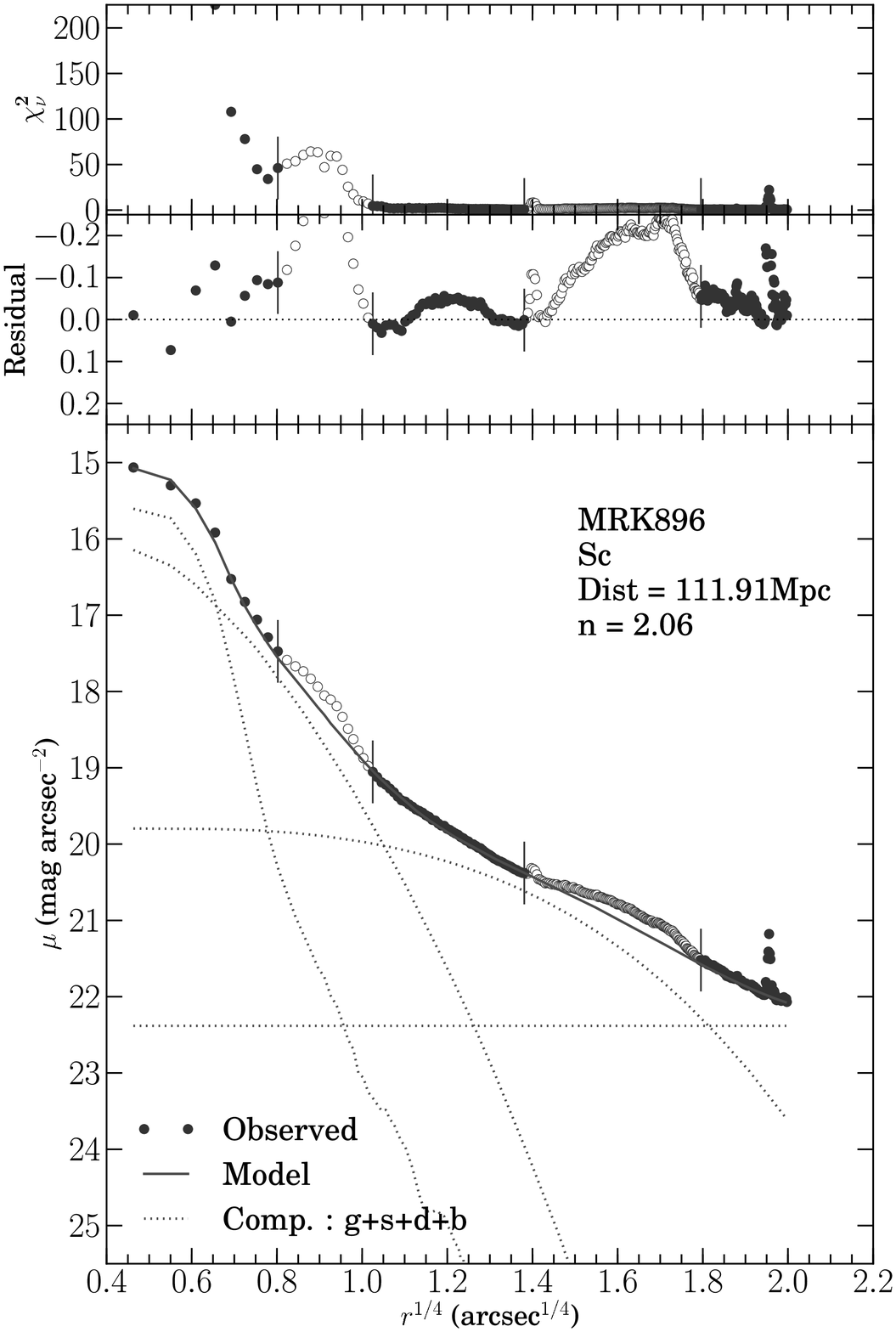}}
     \caption{Bulge-Disk decomposition illustration from the NLS1 sample. The radial ranges of the mask are indicated by open symbols and straight lines in the radial profiles. The upper plot gives the radial distribution of the reduced $\chi^2$, the middle plot is the magnitude difference $\Delta \mu$ between the original and the model images, the bottom plot presents the observed (dots), the total model (solid line), the modelled bulge and the modelled disk (dashed) light  distributions. The radius is given in $r^{1/4}$ to emphasize the central region.  }
     \label{fig:fit-example-nls1}
\end{figure*}

\begin{figure*}
     \centering
     \subfigure{\includegraphics[height=8.5cm]{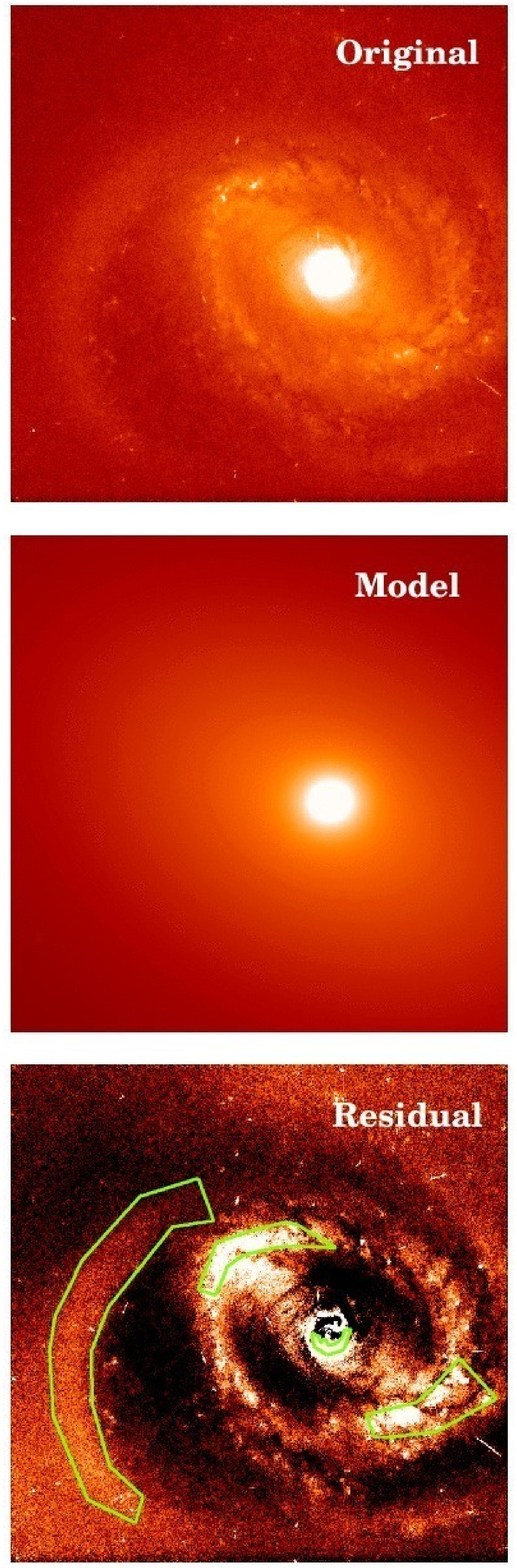}
     \includegraphics[height=8.5cm]{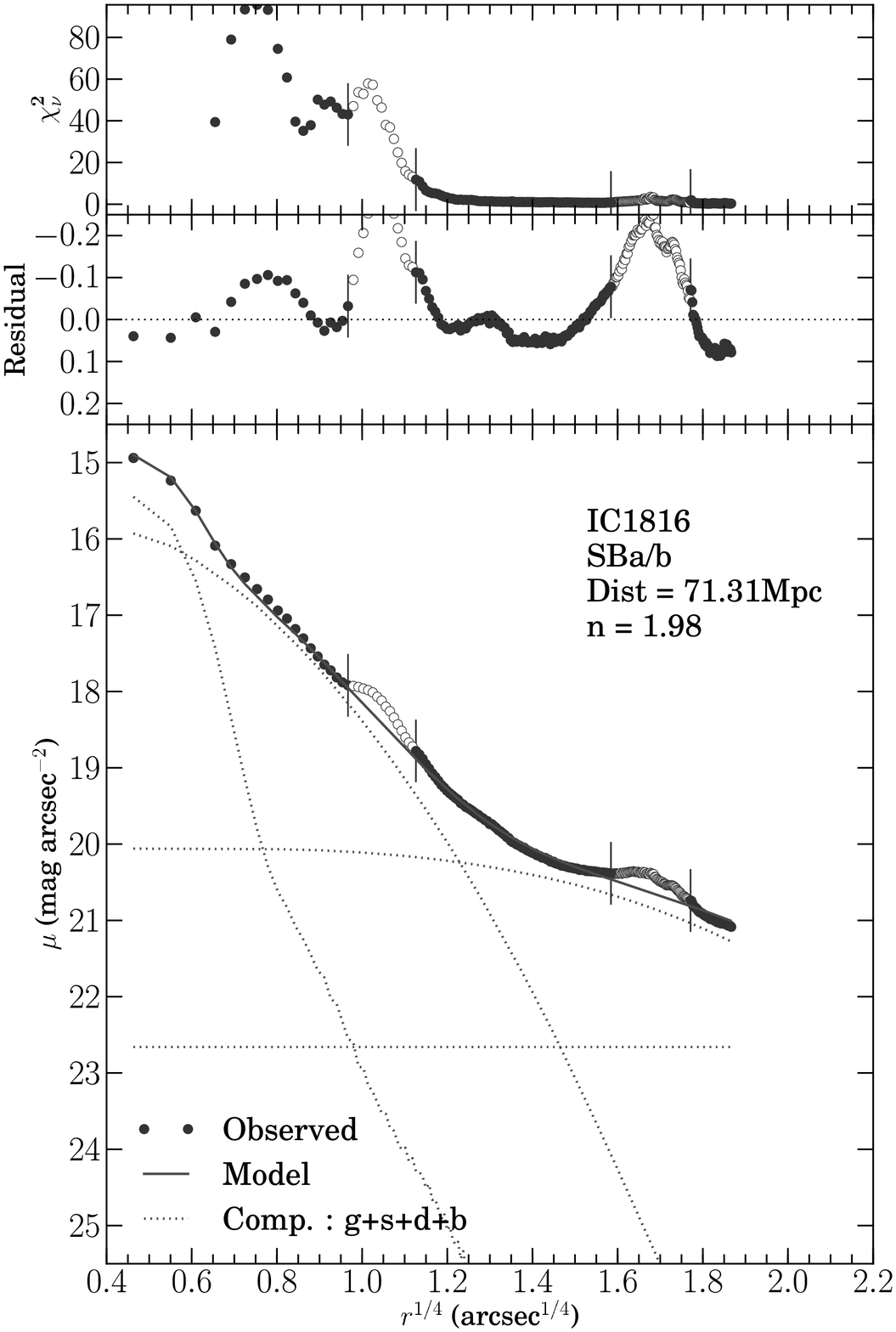}}
     \subfigure{\includegraphics[height=8.5cm]{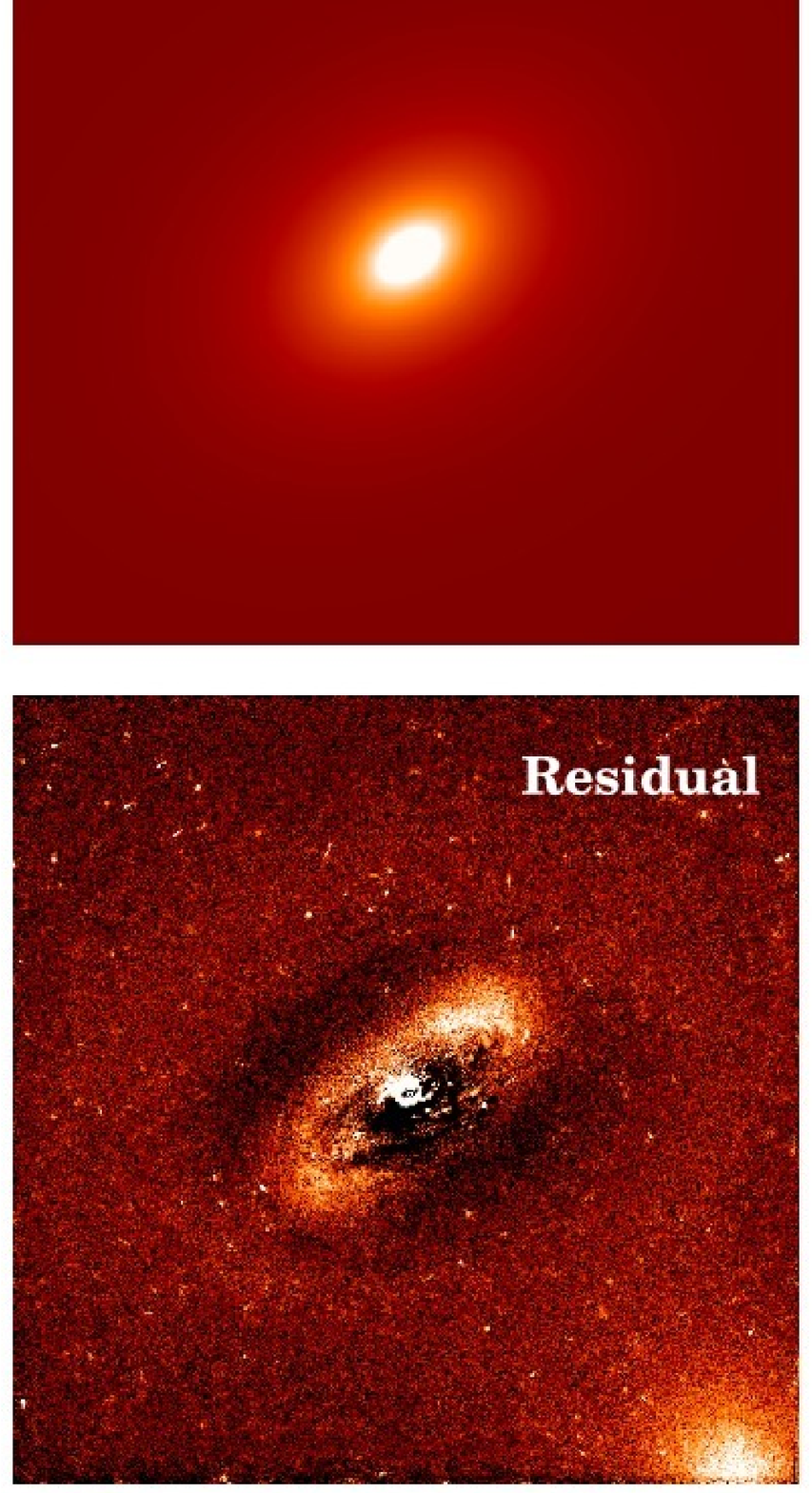}
     \includegraphics[height=8.5cm]{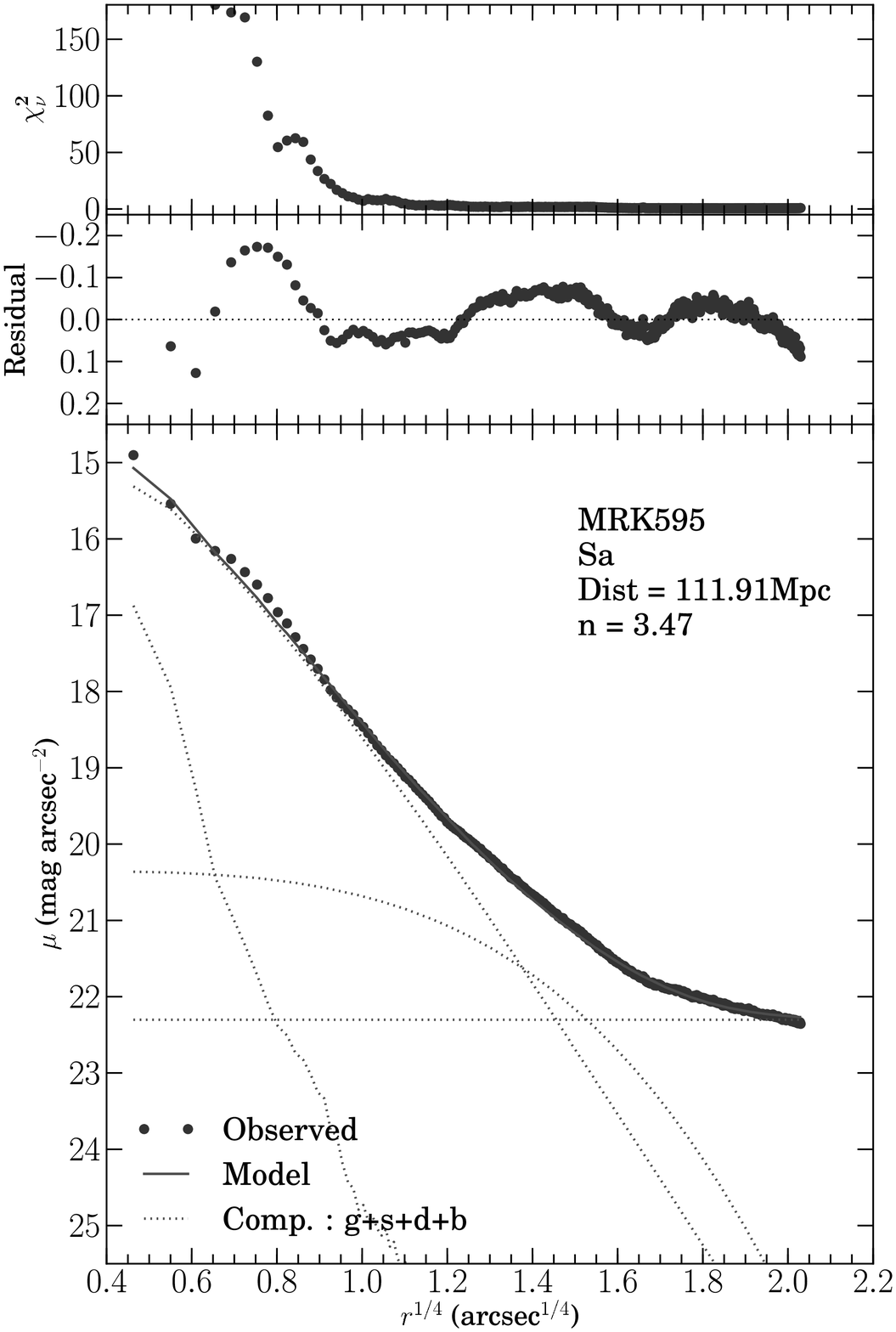}}
     \subfigure{\includegraphics[height=8.5cm]{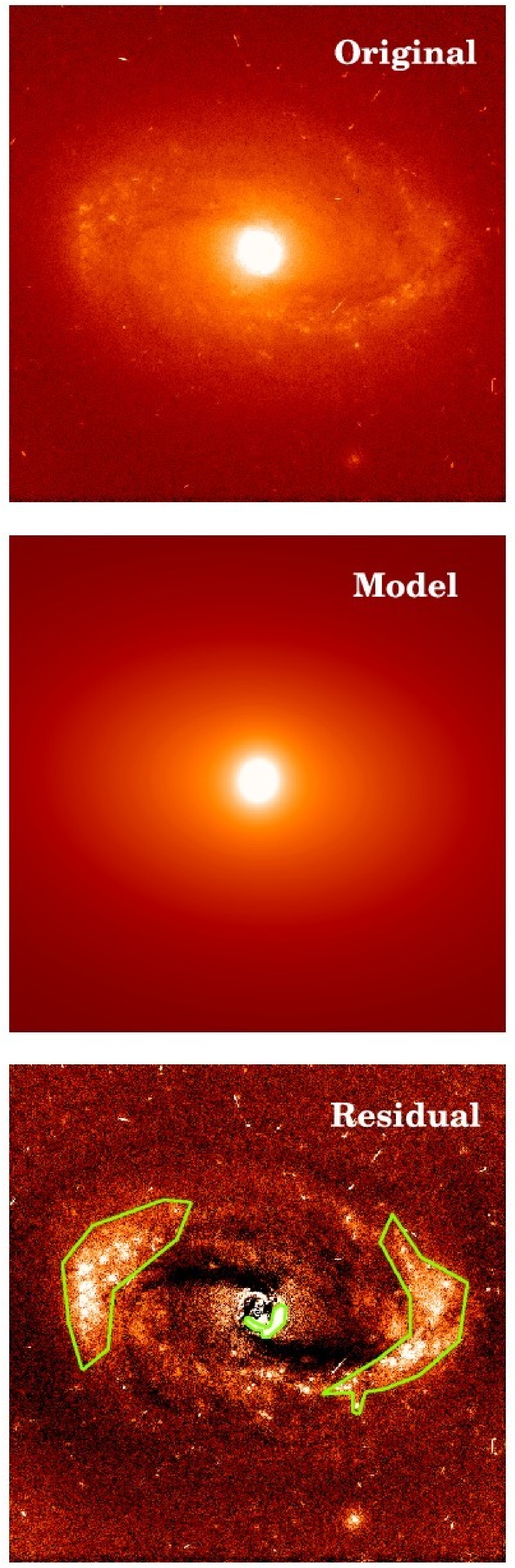}
     \includegraphics[height=8.5cm]{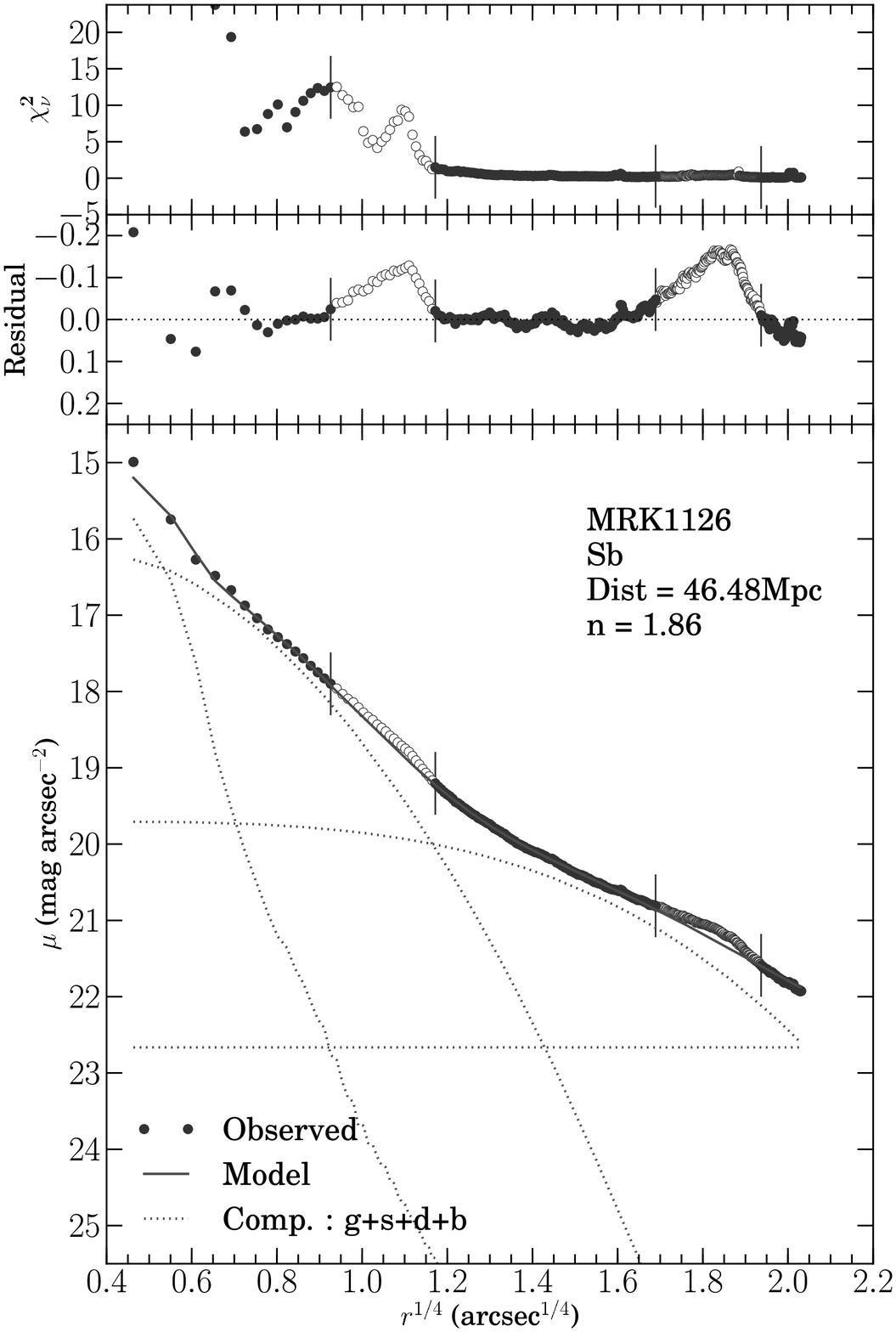}}
     \subfigure{\includegraphics[height=8.5cm]{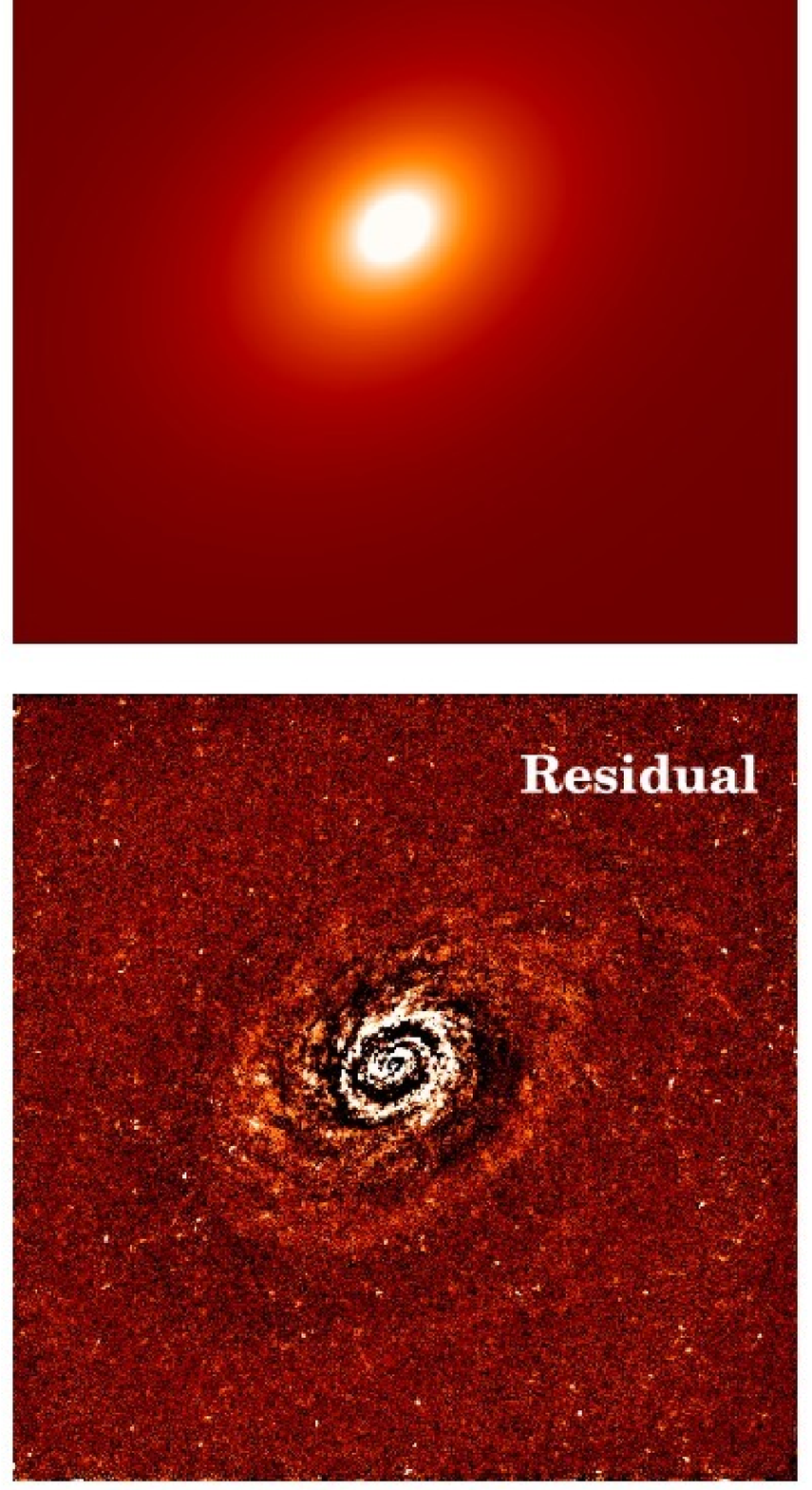}
     \includegraphics[height=8.5cm]{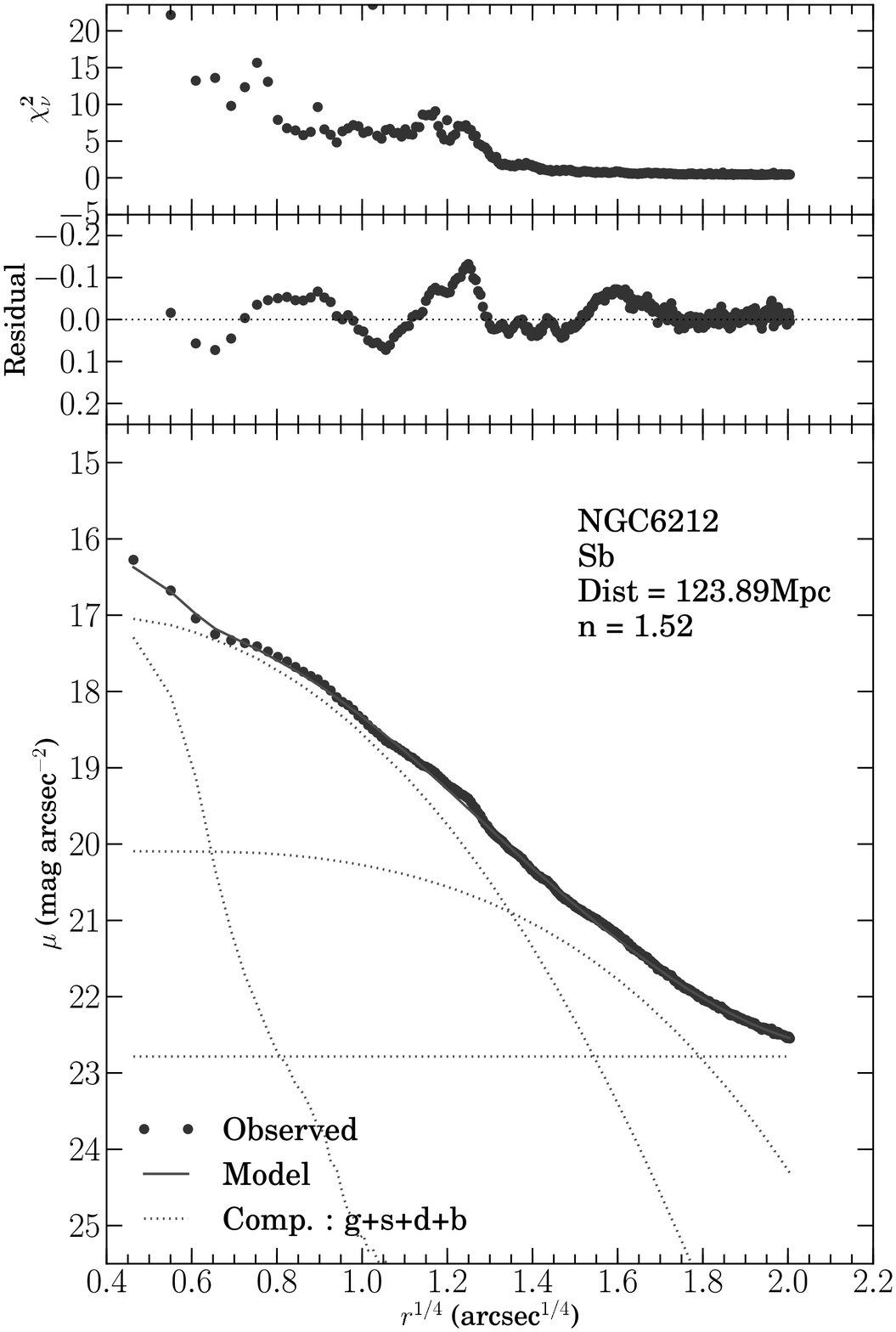}}
     \caption{Bulge-Disk decomposition illustration from the BLS1 sample. The radial ranges of the mask are indicated by open symbols and straight lines in the radial profiles. The upper plot gives the radial distribution of the reduced $\chi^2$, the middle plot is the magnitude difference $\Delta \mu$ between the original and the model images, the bottom plot presents the observed (dots), the total model (solid line), the modelled bulge and the modelled disk (dashed) light  distributions. The radius is given in $r^{1/4}$ to emphasize the central region.  }
     \label{fig:fit-example-bls1}
\end{figure*}

While the \galfit outputs provide errors, those are purely statistical and are insignificant compared to systematic errors.
In the following sections, we discuss the robustness of our fits to different parameters: the treatment of additional structures (deviation from idealized profiles), the unprecise knowledge of the background level, the core saturation and the PSF.
Those parameters give indirect information on the systematic errors.
Finally, we give in  \Fref{fig:fit-example-nls1} and \Fref{fig:fit-example-bls1} eight examples of our fits where we indicate by open symbols the radial ranges entirely or partially masked. The radial residual and the radial $\chi_{\nu}^2$,  given in the upper panels, also provide indirect information on the fit errors.

\subsection{Choice of fitting range: treatment of additional structures} \label{app:fit11}
As already mentioned, any nuclei is accounted for in the fits by the use of a small Gaussian profile. The alternative consists of removing the nucleus from the fit by masking it. Nevertheless, with such a procedure, the bulge-disk decomposition is sensitive to the quality of the mask of the nucleus and can also be affected by the reduced number of constraints, \ie the bulge can  be underconstrained if the mask is too large.
Despite these considerations, we try the alternative and mask systematically the central region of the images. We use a default circular mask with 10 pixels radius, adjusting it only in a few cases (4/28) to have a minimum radius of 5 times the standard deviation $\sigma$ of the Gaussian component (2/4) or to keep a reasonable number of constraints for the bulges (2/4).  The median S\'ersic index difference between the fits with nucleus masked  and the fits with the nucleus included is reasonably small: 0.22 and 0.29 for the NLS1 and BLS1 sample respectively. This test shows that to mask the nucleus instead of fitting it by a Gaussian component biases slightly the S\'ersic indices to higher values. Nevertheless, except for a few cases (3/28 with $\Delta n_b > 2$ and 2/28 with $2>\Delta n_b >1$), the increase remains small, acknowledging the robutness of the fits.

As our samples are made of high resolution HST data, they also present many structural details.
The effect of deviations of surface brightness such as rings, bars or spirals are considered in the last step of our fitting procedure (step 4). Upon examination of the radial profiles and the residual image, we identify, if any, potential additional structures and manually create appropriate 2D masks. We then refit our model and iteratively adjust the mask according to the radial profile and residual image. 
We recognize that this practice is subjective but nonetheless we believe it to be necessary. 
Indeed, modelling structural details -- such as inner rings -- is a complex task, and the fit of any additional component requires more constraints, which cannot be provided by our single snapshot HST images (indeed the current bulge/disk models already give $\chi^2_{\nu} \sim 1$ for most of the fits). The alternative practice of not masking these structures could lead to wrong results. 
For example, in the case of MRK 42, presented in \Fref{fig:fit-example-nls1}, we first do not mask the inner-ring  and obtain a compact bulge with $n_b\sim0.7$ (in our step 3), while after masking the stucture,  we obtain a more reasonable S\'ersic index $n_b\sim1.27$ (in our step  4), which -- according to the radial profile in \Fref{fig:fit-example-nls1} -- is a much more accurate model of the bulge.
Finally, our iterative process ensures that the regions we mask are physical additional structures in the galaxies.

\subsection{Background level}

Given that the background may be coupled to the disk profile, an important aspect of bulge-disk decomposition is to correctly fix the sky level.
Since no precise sky measurements are available for our HST images and  that the galaxies filled most of the field of view, we have to fit the background level together with the other components and minimize the coupling with the disk profile. Therefore, we take a particular care to any additional light profiles in the images such as stars, satellite galaxies or remaining large cosmic rays. These are carefully masked so to minimize their influence on the sky level. Despite this particular attention, we fix the sky level in 2 cases -- MRK279 and MRK609, both part of the BLS1 sample -- at the value obtained in our step 1 (\ie fit of the disk and the background only). If the sky is not fixed, the resulting parameters are not physical: in the  case of MRK 279 the background becomes negative and the radius of the disk excessively large; in the  case of MRK 609 the sky becomes extremely large and the disk and the bulge shrink (with  $n_b < 0.5$).

\section{Reliability of the bulge/disk decomposition} \label{app:fit2}

We analyze here possible effects influencing our fit results~: the core saturation in the images and the choice of PSF. 

As already mentioned, a non-negligible fraction of the images presents saturated core with charges leakage: 50\% in the NLS1 sample and 26\% in the BLS1 sample. To minimize any effect on our fits, we mask carefully the saturated pixels and the pixels affected by charges leakage. Looking at the S\'ersic indices, we find that galaxies with saturated core have a mean S\'ersic index $<n_b> =2.27$ and galaxies without saturated core have $<n_b> =2.06$. Therefore, if the saturation still affect our images, it would tend to increase the S\'ersic index. As our NLS1 sample is more affected by saturation than our BLS1 sample, it would tend to increase the mean S\'ersic index of NLS1s. Consequently, any remaining effect of the saturation cannot account for the difference between NLS1 and BLS1 host bulge S\'ersic indices but would tend to decrease it.

We test the dependence of our fit to the PSF used for convolution by refitting our NLS1 sample with different PSFs. 
The PSFs for this test are also generated with \tinytim but, instead of using an uniform weight along the wavelength range, they are produced at the central mono-wavelength of the filter F606W.
The difference in S\'ersic index ranges from 0 to 0.01 except for one object where it is 0.06. Therefore, we conclude that the PSFs are not critical in our fits and that our choice of using an uniform weight along the wavelength range to create the PSFs with \tinytim is acceptable.

\end{document}